\documentclass[%
 reprint,
 amsmath,amssymb,
 aps,
]{revtex4-1}

\usepackage{graphicx}
\usepackage{dcolumn}
\usepackage{bm}
\usepackage{xcolor}

\begin{document}

\preprint{APS/123-QED}


\title{Spatial Strength Centrality and the Effect of Spatial Embeddings on Network Architecture}

\author{Andrew Liu}
\affiliation{Department of Mathematics, University of Utah, Salt Lake City, UT 84112, USA
}

\author{Mason A. Porter}
\affiliation{Department of Mathematics, University of California, Los Angeles, CA 90095, USA
}

\date{\today}

\begin{abstract}

For many networks, it is useful to think of their nodes as being embedded in a latent space, and such embeddings can affect the probabilities for nodes to be adjacent to each other. In this paper, we extend existing models of synthetic networks to spatial network models by first embedding nodes in Euclidean space and then modifying the models so that progressively longer edges occur with progressively smaller probabilities.
We start by extending a geographical fitness model by employing Gaussian-distributed fitnesses, and we then develop spatial versions of preferential attachment and configuration models. We define a notion of ``spatial strength centrality'' to help characterize how strongly a spatial embedding affects network structure, and we examine spatial strength centrality on a variety of real and synthetic networks.

\end{abstract}

\maketitle




\section{Introduction} \label{sec:introduction}

Many networks have important spatial features, and it is often useful to think of the nodes of such spatial networks as being embedded in either a physical or a latent space \cite{barthelemy,newman2018,geometry-review}. In nature, such a space can be literal, like the physical distance between different parts of a city, or it can be more abstract. In social networks, for example, one can suppose that the nodes that represent individuals have a distance between them that represents physical distance; a distance between them that arises from demographic characteristics, interests, behaviors, or other features; or a combination of such distances \cite{social}. In food webs, nodes that represent individuals may be embedded in a latent space that represents various features and affects their likelihood of interacting with each other. For instance, in a standard niche model, one places each species on a line, with a position that represents its mass and directed edges between nodes to represent who eats whom \cite{foodwebs}.
 
Physical networks, such as transportation networks and other networks within and between cities, possess a natural spatial embedding based on their physical location in the world. Whether networks are embedded in an explicit or latent space (or both), the distances between nodes can strongly influence whether they are adjacent to each other in a network \cite{geometry-review,spatial1,air-traffic,routelengthstatistic,papa2018}. Intuitively, nodes that are farther apart from each other have fewer opportunities to interact (or their interaction has a greater associated cost), so we expect that there is a concomitant lower probability to observe edges between them. For example, people in a social network may be unlikely to interact if they share little in common \cite{socialdistance}, and animals in a food web may not hunt others that are too large or small in comparison to them \cite{foodwebs}. 

It is sensible that spatial characteristics should influence network structure, but it is much more difficult to make these ideas precise \cite{barthelemy}. For example, how exactly does a spatial embedding influence the architecture --- both topology and edge weights --- of a given network? For simplicity, most studies of networks have ignored spatial embeddings, but new models that incorporate spatial considerations are being developed with increasing frequency \cite{barthelemy}. It is also desirable to develop spatial network models that can produce realistic structural features of networks \cite{geometry-review,boguna2019}.

Network models that incorporate space can improve the measures of importance (i.e., centrality) in networks \cite{spatial1, air-traffic,sarkar2019}, as many networks that one constructs from empirical data are affected by space. Spatially relevant features have been examined in examples such as brain networks \cite{braingrowth1, braingrowth2}, fungal networks \cite{fungal_data}, road networks \cite{road_data, spatial1, barbosa, boeing2018multi}, networks of flights between airports \cite{air-traffic}, gas-pipeline networks \cite{spatialefficiency}, granular networks \cite{berthier2019,nauer2019}, social networks \cite{danchev2018,sarkar2019}, and other applications. Incorporating spatial features can improve the modeling of empirical data \cite{barthelemy}, and it is therefore important to further extend ideas from network analysis into the realm of spatial networks.

Examples of spatial models include spatially-embedded random networks \cite{penrose-rgg, geographical_threshold} and networks that grow in time (e.g., through preferential attachment) \cite{mean_field_evolving_spatial, SPA1, spatial1}. Such models can provide reference models with which to compare empirical networks. For example, ideas from spatial networks have led to the development of spatial null models for community detection \cite{community1, community2}. Some spatial network models produce networks with properties that are reminiscent of empirical networks, such as simultaneously having large values of clustering coefficients and heavy-tailed degree distributions \cite{geometric_preferential_attachment, geographical_threshold2}. Spatial network models have also been helpful for studying models of biological growth, such as osteocyte-network formation \cite{mean_field_evolving_spatial} and leaf-venation networks \cite{leaf_optimization}.

In the present paper, we explore a general approach for extending non-spatial network models to spatial versions by introducing a deterrence function \cite{community2}. This function $h(r)$, where $r$ represents distance in either latent or physical space, modifies the probability that nodes are adjacent to each other as a function of the distance between them. We start by examining a modification of a geographical fitness (GF) model \cite{yusuke} that is a latent-space (``hidden variable'') model. Other papers that examined this GF model \cite{geographical_threshold, geographical_threshold2} used exponential and power-law fitness functions, but we use one that has a Gaussian distribution. We then explore a spatial extension for the Bar\'abasi--Albert (BA) preferential-attachment model, where the deterrence function $h(r)$ modifies the attachment probability for new edges. This too has been examined in earlier papers \cite{SPA1, SPA2, SPA3}, but we incorporate a deterrence function in a slightly different way. Finally, we apply the deterrence function $h(r)$ to a configuration model \cite{fosdick} to create a spatial configuration model. We compute some characteristics of our spatial network models to understand how a deterrence function affects network structure. We also define a new spatial notion of centrality, which we call ``spatial strength centrality'', that helps us measure how strongly a spatial embedding affects the topological structure of a network. We examine how spatial strength centrality behaves on our new synthetic models, and we also compute it for a variety of empirical spatial networks.

Our paper proceeds as follows. In Section \ref{sec:fitness_model}, we discuss GF models and explore the properties of these networks when using Gaussian-distributed fitnesses. We study a spatial preferential attachment (SPA) model in Section \ref{sec:ba-model} and explore its properties, and we develop a new spatial configuration model in Section \ref{sec:configuration_model}. In Section \ref{sec:spatial_strength}, we define our notion of spatial strength centrality and apply it to several empirical and synthetic networks, including the models that we explored in previous sections. We conclude in Section \ref{sec:discussion}.


\section{Geographical Fitness Model} \label{sec:fitness_model}

\subsection{Prior research}

The (non-geographical) fitness network model is a class of networks that assigns fitness values to the $n$ nodes of a network \cite{yusuke}. Such models are sometimes also called ``threshold models'' \cite{geographical_threshold}, although one needs to be careful not to confuse them with other similarly-named models \cite{porter2016dynamical,newman2018}. One determines the intrinsic fitnesses of the nodes from a density function $f(w)$. In the original threshold model of this type \cite{caldarelli}, the intrinsic fitness characterizes the propensity of a node to gain edges. Such models have been used to generate small-world networks with power-law degree distributions \cite{geographical_threshold}.

One choice of fitness distribution is the exponential distribution
\begin{equation}\label{exponentialfitness}
    f(w) = \lambda e^{-\lambda w}\,, \quad w \geq 0\,.
\end{equation}
It gives the probability that a node has an intrinsic fitness value of $w$, where the parameter $\lambda \geq 0$ determines the shape of the distribution. An edge exists between nodes $v_i$ and $v_j$, with $i \neq j$, when 
\begin{equation} \label{chosen}
    g(i,j) = w_i + w_j \geq \theta \,,
\end{equation}
where $\theta$ is a ``threshold parameter'' that determines the fitness values that nodes need to be adjacent to each other. With equation \eqref{chosen}, nodes that have a higher fitness $w$ also have a larger degree. With either an exponential distribution or power-law distribution of fitness, one can show that the resulting degree distribution follows the power law $p(k) \sim k^{-2}$ as $k \rightarrow \infty$ \cite{caldarelli, threshold}.

One can also formulate geographical (or, more generally, spatial) versions of a fitness model (so-called ``GF models''), as was illustrated in \cite{geographical_threshold, boguna, caldarelli}. In one extension, a network exists in a $d$-dimensional Euclidean space of finite size, such as $[0, 1] \times [0, 1] \subset \mathbb{R}^2$ for $d = 2$. In addition to assigning node fitnesses using the distribution $f(w)$, one now also assigns a location in space to each node $v_i$. For example, perhaps one determines each of the node's $d$ coordinates uniformly at random in the space. One can then suppose that an edge exists between nodes $v_i$ and $v_j$, with $i \neq j$, when
\begin{equation}
    g(i, j, r) = (w_i + w_j)h(r) \geq \theta \,,
\end{equation}
where $r$ is the Euclidean distance between nodes $v_i$ and $v_j$, and the function $h(r)$ describes the influence of space on the connection probability of two nodes as a function of the distance between them. The function $h(r)$ is sometimes called a ``deterrence function'' \cite{barbosa}, and it has been used in many studies of spatial networks \cite{barthelemy}. A common choice for a deterrence function is the power-law form \cite{geographical_threshold}
\begin{equation}\label{distance_equation}
    h(r) = r^{-\beta}\,,
\end{equation}
for some value of a spatial decay parameter $\beta$. For certain values of $\beta$, it is possible to determine exact expressions for degree distributions and global clustering coefficients for the associated fitness-network model \cite{geographical_threshold}, but most investigations with this deterrence function have focused on numerical computations.


\subsection{Gaussian-distributed fitnesses}

In previous treatments of (geographical and non-geographical) fitness network models, it has been very common to choose exponential and power-law distributions of fitnesses, including for the specific purpose of generating networks with power-law degree distributions \cite{geographical_threshold, geographical_threshold2, caldarelli, boguna}.

In the present study, we use a Gaussian distribution for fitness, rather than an exponential or power-law distribution. We make this choice based on the intuition that entities (which are represented by nodes) have a variety of intrinsic factors that influence whether they interact with other nodes in a system. We model the probability than an edge exists between two nodes as a function of a single fitness parameter. By assuming that fitness is an aggregation of many intrinsic factors (in which randomness plays a role), it seems reasonable to assume that it has a Gaussian distribution \cite{frank}. Specifically, we employ the standard normal distribution
\begin{equation}
        W {\raise.17ex\hbox{$\scriptstyle\mathtt{\sim}$}} N(0, 1)\,.
\end{equation}
We also modify the threshold function and write
\begin{equation}
    g(i, j, r) = |w_i - w_j| h(r) \geq \theta\,,
\end{equation}
such that nodes that differ widely in intrinsic fitness are more likely to interact with each other than those with similar intrinsic fitnesses. That is, we generate networks with disassortative mixing with respect to node fitness. Examples where this may be relevant include hyperlinks from many low-fitness Web pages to high-fitness ones and predators in a food web who hunt much weaker prey.


\subsection{Network realizations and numerical computations}\label{sec:fitness_numerics}

For our numerical computations, we construct networks with $n = 500$ nodes that are embedded in a box $[0, 1] \times [0, 1]$ with periodic boundary conditions. The nodes have coordinates $(x_1, x_2)$, where we assign $x_1$ and $x_2$ independently with uniform probability on the interval $[0,1]$, and Gaussian-distributed fitnesses with mean $\mu = 0$ and standard deviation $\sigma = 1$. 

For the threshold function \eqref{distance_equation}, we consider values $\beta \in \{ 0, 0.5, 1, 1.5, 2, 2.5, 3 \}$ for the decay parameter $\beta$, and we manually adjust $\theta$ such that the mean degree is $\langle k \rangle = 20$. We obtain $\theta$ through a combination of trial-and-error and fitting through linear regression. We generate $30$ instances of these networks in this GF model for each value of $\beta$.

\begin{figure*}
    \centering
    \includegraphics[width=0.7\linewidth]{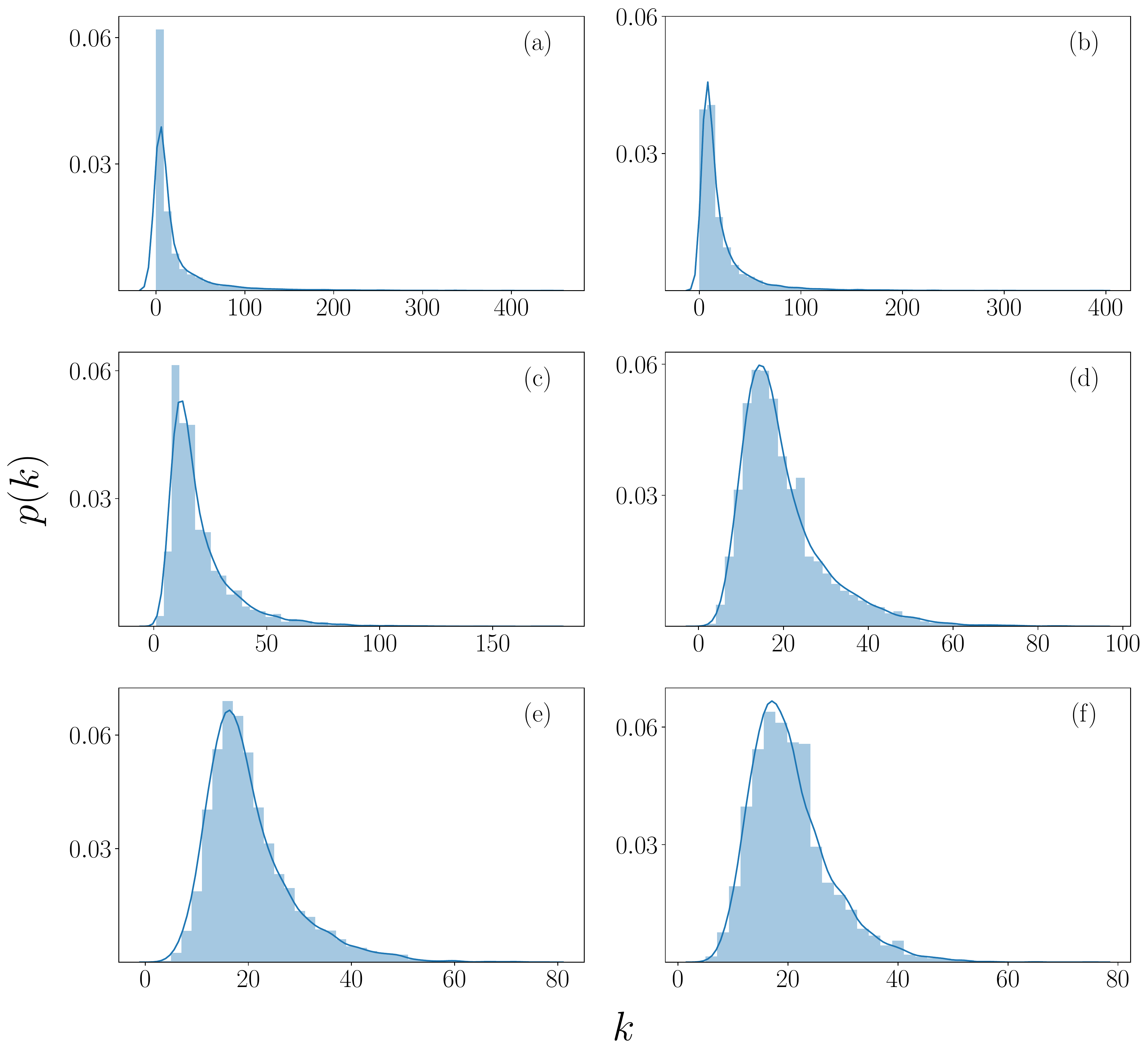}
    \caption{Example degree distributions of geographical fitness networks with Gaussian-distributed fitnesses and various values of the decay parameter $\beta$. The values of $\beta$ for each of the panels are (a) 0 (b) 0.5, (c) 1, (d) 1.5, (e) 2, and (f) 2.5.
    }
\end{figure*}

When $\beta = 0$, we recover a non-geographical threshold model, where the distances between nodes do not play a role in the probability for nodes to be adjacent to each other. In this regime, the distribution looks visually like it may satisfy a power law, although we do not test this idea. For progressively larger values of $\beta$, distance plays a progressively stronger role, and nodes that are farther apart are less likely to be adjacent to each other.

We calculate a few well-known network characteristics \cite{newman2018} and find that they are affected by distance (as encoded in the value of $\beta$). The characteristics that we calculate are the following ones.
\begin{enumerate}\label{characteristics_definitions}
    \item {\bf Mean local clustering coefficient.} The local clustering coefficient of a node with degree at least $2$ is $c_i = \frac{2T(v_i)}{k_i(k_i - 1)}$, where $T(v_i)$ is the number of unique triangles (i.e., $3$-cliques) to which node $v_i$ belongs and $k_i$ is the degree of node $v_i$. A node with degree $0$ or $1$ has a local clustering coefficient of $0$. The mean local clustering coefficient of a network is the mean value of $c_i$ over all nodes in the network (including those with degrees $0$ and $1$). 
    \item {\bf Mean geodesic distance.} The mean geodesic distance is  $L = \frac{\sum_{i \neq j}l(v_i, v_j)}{n(n-1)}$, where $l(v_i, v_j)$ is the shortest-path distance (in terms of number of edges) between nodes $v_i$ and $v_j$.
    \item {\bf Mean edge length.} We take the length of an edge to be the Euclidean distance between its two incident nodes. The mean edge length of a network is the mean of this length over all edges in the network.
    
    \item {\bf Degree assortativity.} We calculate the degree assortativity of a network using the Pearson correlation coefficient $r = \frac{\sum_{i,j}(A_{i,j}-k_i k_j/2m)k_i k_j}{\sum_{i,j}(k_i \delta_{i,j}-k_i k_j/2m)k_i k_j}$ \cite{newman2018}, the normalized covariance of degrees of the nodes of the network.
\end{enumerate}

Larger values of $\beta$ yield shorter mean edge lengths, larger mean geodesic distances, and larger mean local clustering coefficients. In our plots in this section, we generate 30 instantiations of the geographical fitness model for each value of $\beta$. We then compute these characteristics for each network and plot their means.

Our result for the mean local clustering coefficient is intuitively sensible. For progressively larger values of $\beta$, spatial effects become more prominent, and spatially close nodes become more likely to be adjacent to each other. As an example, consider a pair of nodes, $v_1$ and $v_2$, that are both adjacent to node $v_0$. For progressively larger values of $\beta$, $v_1$ and $v_2$ are likely to be progressively closer to each other, because the edges $(v_0, v_1)$ and $(v_0, v_2)$ are progressively more likely to be shorter. Therefore, the edge $(v_1, v_2)$ has a larger probability to exist. This intuition suggests that the mean local clustering coefficient should increase with $\beta$.
 
To facilitate exposition, we use the term ``spatial effects'' to describe the influence of a network's spatial embedding on its topology and network characteristics, such that ``greater'' spatial effects signify more influence of a spatial embedding on a network. With the deterrence function $h(r) = r^{-\beta}$, we expect spatial effects to increase as we increase $\beta$. The most obvious example of this phenomenon is with the mean edge length of a network. With this model (and the subsequent ones in this paper), as we increase $\beta$, edges form increasingly preferentially between spatially close nodes, so mean edge length decreases as we increase $\beta$. The aforementioned increase in mean local clustering coefficient with $\beta$ is also a ``spatial effect''.

\begin{figure}
    \centering
    \includegraphics[width=1.0\linewidth]{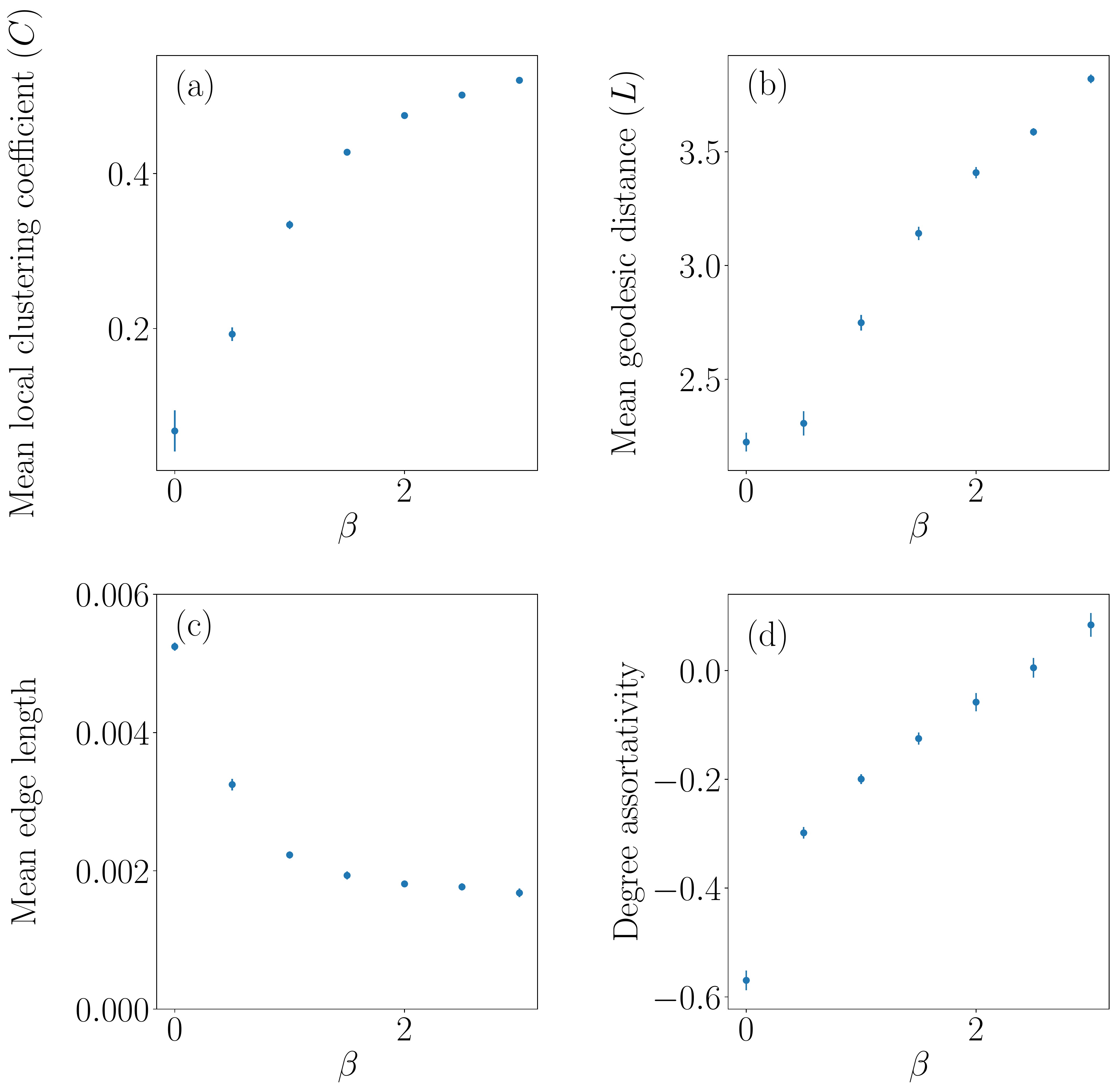}
    \caption{Characteristics of networks (with $n = 500$ nodes) from a geographical fitness model with Gaussian-distributed fitness for different values of the spatial decay parameter $\beta$ and $30$ instantiations of the model for each value of $\beta$. We show the (a) mean local clustering coefficient, (b) mean geodesic distance, (c) mean edge length, and (d) degree assortativity. In all cases, we first take means with respect to individual networks and then take means over the $30$ instantiations. The error bars indicate 95\% confidence intervals.
    }
    \label{fig:gf_metrics}
\end{figure}

Near $\beta = 0$, we observe a small degree assortativity because of the heterogeneous mixing of nodes with different degrees. This is a consequence of how we formulated the geographical fitness model, as nodes with extreme fitness values (either positive or negative ones) yield larger values of $|w_i - w_j|$, so they are adjacent to all nodes with fitnesses that are sufficiently close to $0$. These nodes, whose fitnesses are in the center of the fitness distribution, are not likely to be adjacent to each other.

By contrast, for larger values of the decay parameter $\beta$ (e.g., for $\beta \geq 2$), the Euclidean distance between two nodes has more influence over the probability that they are adjacent. In this regime, it is no longer the case that nodes with extreme fitness values are adjacent to all nodes in the network. If we let $\beta \rightarrow \infty$, we expect the distance between nodes to become the sole factor that determines whether nodes are adjacent. In our numerical computations, we hold $\langle k \rangle$ roughly constant at $20$, so we expect nodes to be adjacent to roughly $20$ of their nearest neighbors. In this respect, as $\beta \rightarrow \infty$, we expect the network to resemble a random geometric graph (RGG) \cite{penrose-rgg, rgg} with an appropriate value of a connection radius $r_c$. We consider the simplest type of RGG, which has two parameters: the number $n$ of nodes and a connection radius $r_c$. We add each node to the region $[0, 1] \times [0, 1]$, and we assign its location uniformly at random in the space. Pairs of nodes are adjacent to each other if their Euclidean distance is less than or equal to $r_c$, so adjacency in the network is determined completely by the positions of the nodes. 

{To see how the GF model becomes similar to an RGG as $\beta \rightarrow \infty$, we perform the following experiment.} Starting with an instantiation of the GF model with $\beta=50$, we construct an RGG (with periodic boundary conditions) with the same node positions and with $r_c \approx 11.3$ (which implies that $\langle k \rangle \approx 20$). This network has $98\%$ of the same edges as the GF network. The same experiment with $\beta=3$ yields $77\%$ of the same edges, and one with $\beta=1$ yields $52\%$ of the same edges. Random geometric graphs (of the type that we consider) have positive degree assortativity \cite{rgg_correlations}, so for progressively larger $\beta$, we expect degree assortativity to increase to positive values that resemble those of the RGG. As we show in Fig.~\ref{fig:gf_metrics}, this indeed appears to be the case.


\subsection{Closeness centrality}\label{close}

Normalized closeness centrality gives an idea of how ``close'' a node is to other nodes in a network, as determined by how many edges are between it and other nodes in the network \cite{newman2018}. The version of closeness centrality that we calculate is
\begin{equation}
    C(v_i) = \frac{n - 1}{\sum\limits_{i \neq j} l(v_i,v_j)}\,,
\end{equation}
where $l(v_i, v_j)$ is the geodesic distance between nodes $v_i$ and $v_j$.

\begin{figure}
    \centering
    \includegraphics[width=0.8\linewidth]{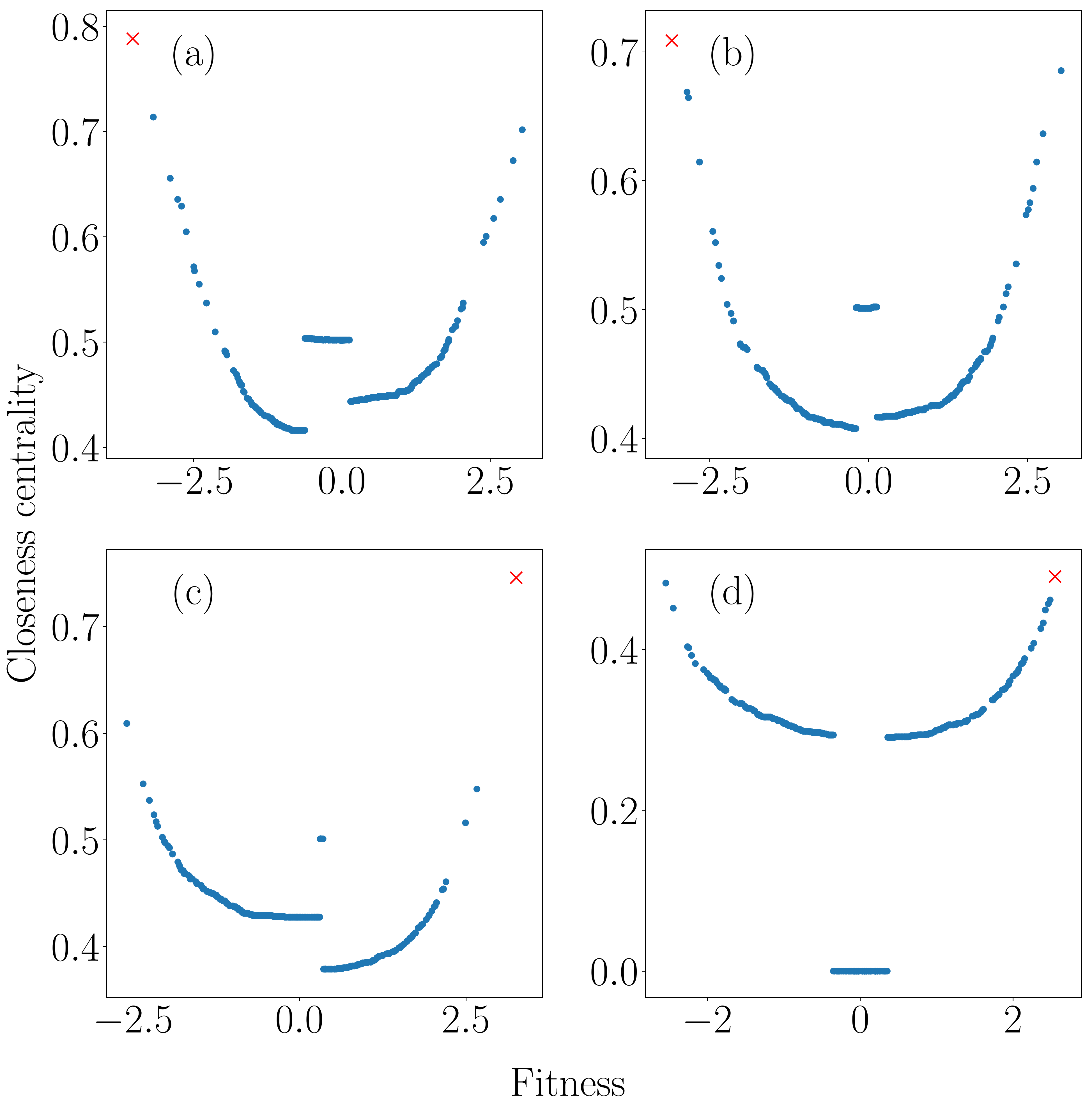}
    \caption{Scatter plot of closeness centralities for nodes in $4$ instantiations of the GF model with decay parameter $\beta = 0$. All instantiations for $\beta = 0$ have similar scatter plots. The depicted instantiations illustrate common patterns. For each network, we highlight the node with the largest (in absolute value) fitness. In most scatter plots, we observe two large branches (on the left and right) and a small horizontal (or predominantly horizontal) curve between these two large branches. We refer to one branch as ``lower'' than the other if the lowest point of that branch is lower than the lowest point of the other branch. We observe that the left branch is lower when there is an ``extremum'' (i.e., a node with the largest {absolute} value) of closeness centrality on the left [see panels (a)--(b)] and that the right branch is lower when there is an extremum of closeness centrality on the right [see panel (c)]. 
    }
    \label{fig:closeness_example}
\end{figure}

\begin{figure}
    \centering
    \includegraphics[width=0.8\linewidth]{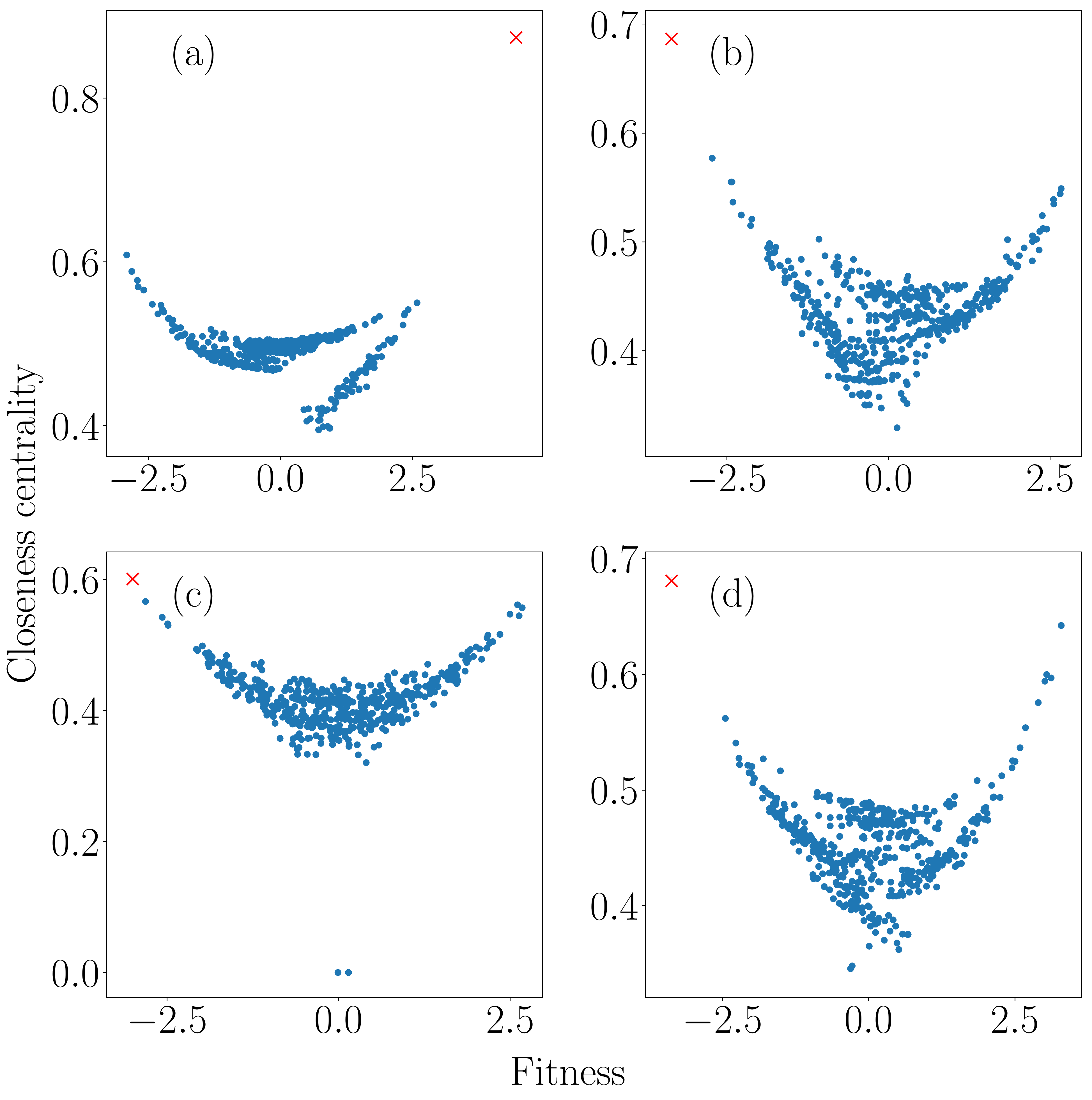}
    \caption{Scatter plot of closeness centralities for nodes in $4$ instantiations of the GF model with decay parameter $\beta = 0.5$. The depicted instantiations illustrate common patterns. For each network, we highlight the node with the largest (in absolute value) fitness. For this value of $\beta$, we do not always observe noticeable branches [see panels (b)--(d)]. When there are noticeable branches in a scatter plot, they are usually accompanied by a node whose fitness has a large absolute value [see panel (a)].
    }
    \label{fig:closeness_example_2}
\end{figure}

\begin{figure}
    \centering
    \includegraphics[width=0.8\linewidth]{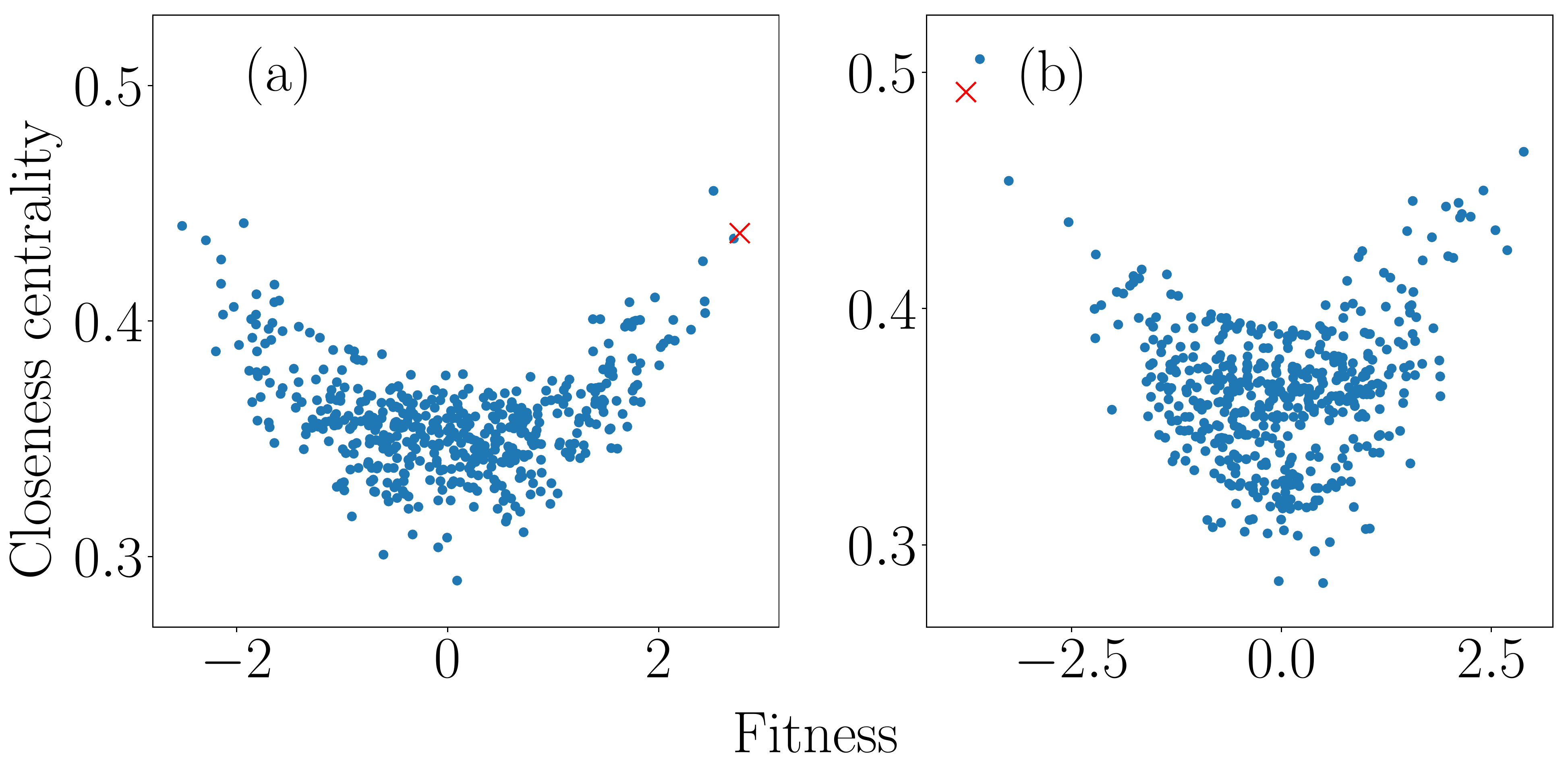}
    \caption{Scatter plot of closeness centralities for nodes in $2$ instantiations of the GF model with decay parameter $\beta = 1$. The depicted instantiations illustrate common patterns. For each network, we highlight the node with the largest (in absolute value) fitness. For this value of $\beta$, branches are almost never apparent in the scatter plots, even with nodes with extreme fitness values [see panel (b)], and the node with the most extreme fitness value is no longer always the one with the largest value of closeness centrality.
    }
    \label{fig:closeness_example_3}
\end{figure}

For $\beta = 0$, the closeness centralities of the nodes arise only from their intrinsic fitness values, as the GF model does not incorporate any spatial effects in this case. In the scatter plots of fitness versus closeness centrality, we observe two large branches of values (see Fig.~\ref{fig:closeness_example}) in all networks from our $30$ instantiations of the model. In these networks, the node whose fitness has the largest absolute value (which we call the ``extremum'') is adjacent to all nodes in the network that have sufficiently different fitnesses; these latter nodes all belong to the other large branch, and the extremum acts as a shortcut to these nodes.

In these scatter plots, we also observe a small central curve between the two branches. It lies below the lowest point of the two branches [such as in Fig.~\ref{fig:closeness_example}(d)] or above the nearby parts of both branches. This curve is below the branches when no nodes in the network have fitnesses with large enough absolute values, such that the nodes in the curve are not adjacent to any other node. Specifically, if the node whose fitness is closest to $0$ has a fitness value of $w_m$ and the extremum has a fitness value of $w_M$, then the small central curve is below the branches when $|w_M| < \theta - |w_m|$. (Note that $\theta \approx 2.9$ for $\beta = 0$.) Otherwise, nodes in the middle curve are adjacent to nodes with very negative or very positive fitnesses. 

Extremum nodes still behave as shortcuts for $\beta = 0.5$, but the spatial constraints necessitate spatial proximity for nodes to be adjacent to other nodes, so the role of the extremum as a shortcut is less prominent. For example, some shortest paths go through other nodes. We show example scatter plots of closeness versus fitness for $\beta = 0.5$ in Fig.~\ref{fig:closeness_example_2}. In Fig.~\ref{fig:closeness_example_2}(a), for example, the presence of an extremum whose fitness has a very large magnitude still yields some branches in the scatter plot. However, in the other examples in this figure (and in most scatter plots from the $30$ instantiations of the GF model), it is more difficult to discern clear branches. For $\beta = 1$ (and hence for larger values of $\beta$), branches are almost never apparent, as node fitnesses are less likely to determine whether two nodes are adjacent; instead, spatial effects become more prominent. We thereby see how spatial effects manifest in this GF model.


\section{A Spatial Preferential-Attachment Model} \label{sec:ba-model}

\subsection{Prior research} \label{prior}

We now examine characteristics of a spatial generalization of the BA preferential-attachment model \cite{BA}. Such a spatial preferential-attachment model (SPA) was first introduced in \cite{SPA1} and explored further in \cite{SPA2, SPA3, SPA4}. It starts with a seed network, with nodes located in $[0, 1] \times [0, 1]$ (using periodic boundary conditions, such that $[0, 1] \times [0, 1]$ is a torus), at $t=0$. During each subsequent time step, one adds a new node and assigns its location uniformly at random in the space. If a node is assigned the same location as an existing node, we then assign a different location uniformly at random. New nodes link to $m$ existing nodes with probability 
\begin{equation}\label{spatialpreferential_prior}
    p(v_i,v_j) = \frac{k_j h(r_{i, j})}{\sum_l k_l h(r_{i,l})}\,,
\end{equation}
where $r_{i,j}$ is the Euclidean distance between nodes $v_i$ and $v_j$, and $h(r_{i,j})$ is the same as in Eq.~\eqref{distance_equation} (i.e., $h(r_{i,j}) = r_{i,j}^{-\beta}$). In \cite{SPA1}, this SPA was simulated on a 1D space with periodic boundary conditions.

References \cite{SPA2, SPA3} examined this SPA model embedded in 2D. Manna and Sen \cite{SPA2} assigned one new edge per incoming node and explored the distribution of edge lengths and the expected degree of a node that is born at time $t$ in a region $[0, 1] \times [0, 1]$ with periodic boundary conditions. Yook et al. \cite{SPA3} attempted to fit empirical internet network data using this SPA model with nodes placed at their geographical location on a world map.

Other spatial generalizations of the BA model have also been studied. In \cite{aiello, emmanuel}, for example, each node has a ``sphere of influence'' with a radius that is proportional to the node's in-degree. Upon the addition of a new node $v_t$, if its location is inside the sphere of influence for an existing node, then $v_t$ has a probability $p$ to create an edge to that node. As $t \rightarrow \infty$, the model exhibits a power-law degree distribution and has a positive value for the mean local clustering coefficient \cite{emmanuel}.


\subsection{Description of our SPA model}

In contrast to the SPA models in \cite{SPA1,SPA2,SPA3}, which used a connection probability that is given by Eq.~\eqref{spatialpreferential_prior}, we instead use a connection probability function of
\begin{equation} \label{spatialpreferential}
    p(v_i,v_j) = \frac{k_j}{\sum_l k_l}h(r_{i,j})\,.
\end{equation}
The intuition behind our choice is that a node's ``popularity'' (which we quantify by its relative degree in a network) is independent of its distance to another node. This allows us to transform an existing non-spatial PA model (such as the BA model) into a spatial variant by multiplying the probability of edge formation by a deterrence function $h(r)$. The probability that an incoming node $v_t$ at time step $t$ attaches to any other node in the network (after normalizing this probability such that it sums to $1$) is thus
\begin{equation} \label{spatialpreferential_normalized}
	p(v_t, v_j) = \frac{p(v_t, v_j)}{\sum\limits_{\substack{j < t \\ \{j  |  v_j \not\in N(v_t)\}}} p(v_t, v_j)}\,,
\end{equation}
where $N(v_t)$ is the set of nodes to which $v_t$ is adjacent (i.e., its neighborhood). When implemented in an algorithm, we assign edges one at a time. Additionally, we do not allow self-edges or multi-edges in our model. 

We assign a location uniformly at random in $[0, 1] \times [0, 1]$ to each node as we add it to the network. We start our network with a seed that consists of a $10$-clique, where each node in this clique is located at a position in $[0, 1] \times [0, 1]$ that we assign uniformly at random. At each time step, we add a new node $v_i$ with $m=5$ stubs (i.e., ends of edges), and we link each of these stubs to an existing node in the network with a probability equal to \eqref{spatialpreferential_normalized}. Nodes with larger degrees are more likely to accrue more edges. When $\beta > 0$, incoming nodes are more likely to connect to nearby nodes than to ones that are farther away. All edges are undirected and unweighted. 

For our computations, we consider decay parameters of $\beta \in \{ 0, 0.5, 1, 1.5, 2, 2.5, 3, 3.5, 4\}$, and we examine $10$ instantiations of our SPA model for each value of $\beta$. We simulate each instantiation of our model for $T$ time steps. We use $T \in \{300, 1000, 3000, 10000\}$, which have $n \in \{310, 1010, 3010, 10010\}$ nodes, respectively, because there are $10$ nodes in the seed network). Each node adds $5$ edges to a network, so $\langle k \rangle \rightarrow 10$ as $T \rightarrow \infty$.


\subsection{Computational results}

\begin{figure}
    \centering
    \includegraphics[width=1.0\linewidth]{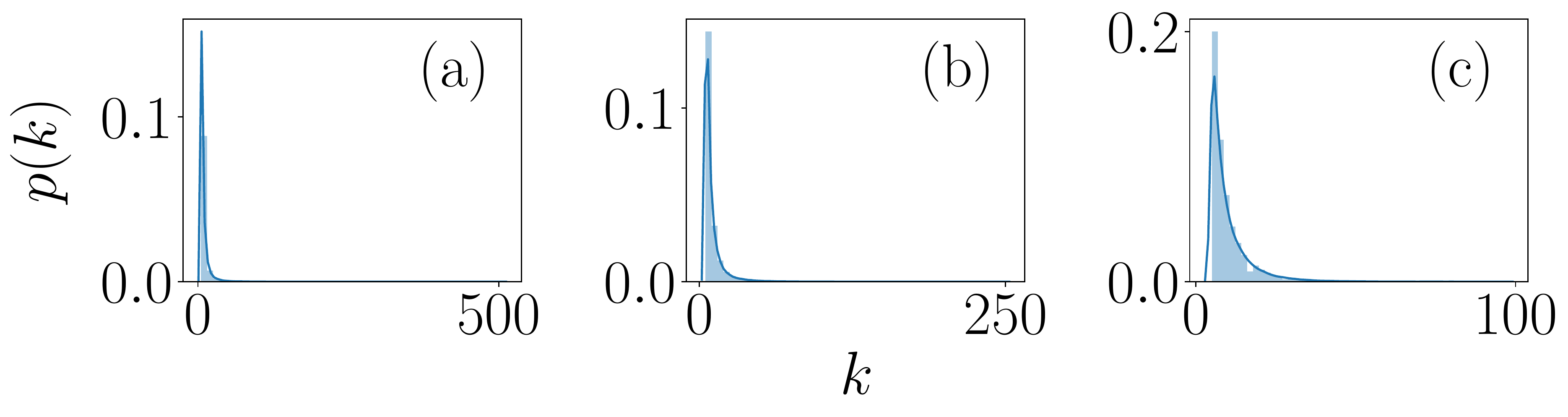}
    \caption{Degree distributions for single instances of our SPA model for $T=10000$ time steps. The values of $\beta$ for each of the panels are (a) 0, (b) 2, and (c) 4.
    }
\end{figure}

\begin{figure}
    \centering
    \includegraphics[width=1.0\linewidth]{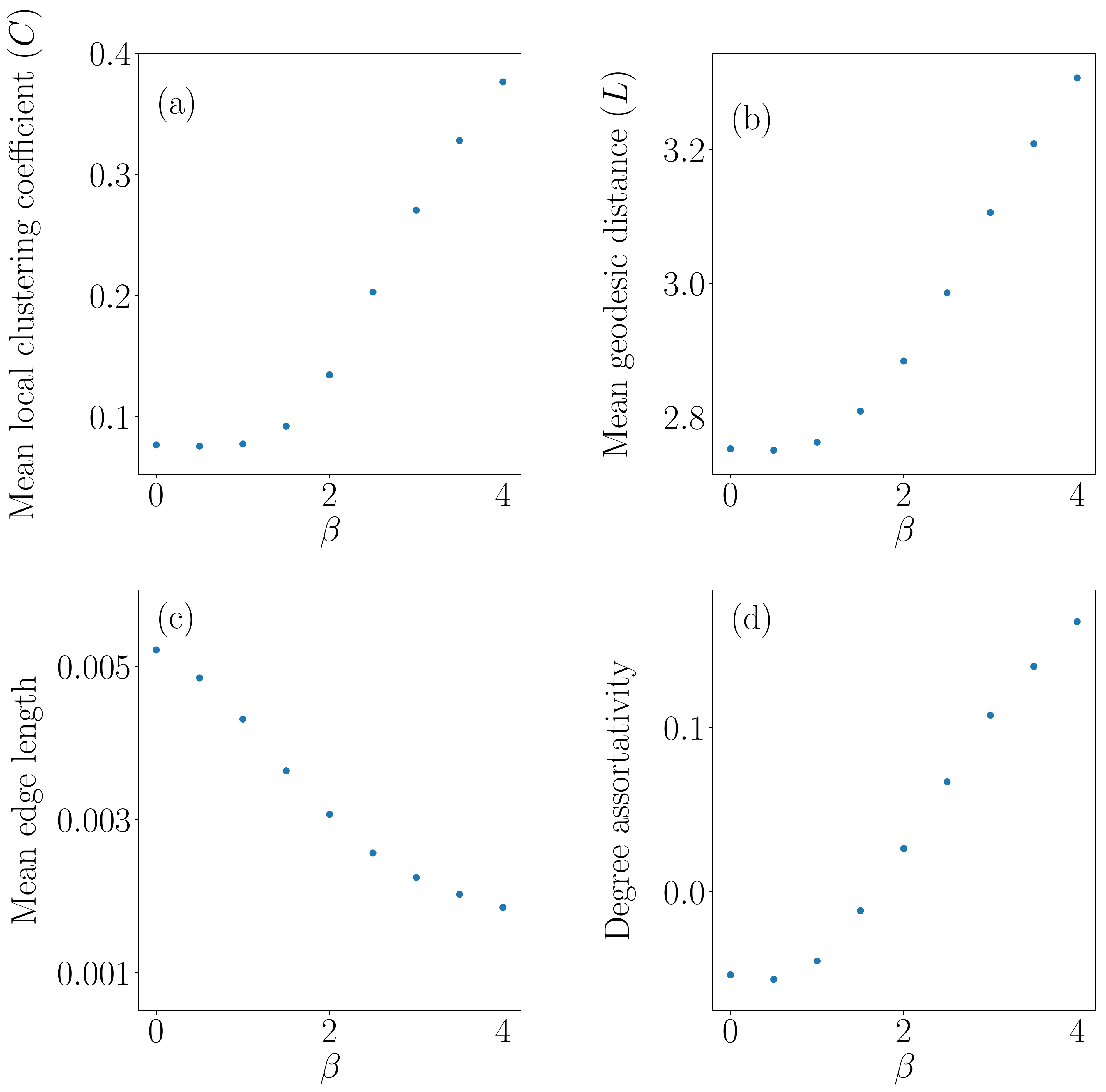}
    \caption{Some characteristics of the networks from our SPA model for different values of the spatial decay parameter $\beta$ for networks with $n=10010$ nodes, $m=5$ edges for each new node as we add to a network, and $10$ instantiations of our model. We show computations of (a) mean local clustering coefficient, (b) mean geodesic distance, (c) mean edge length, and (d) degree assortativity.
    }
\end{figure}

\begin{figure}
    \centering
    \includegraphics[width=0.75\linewidth]{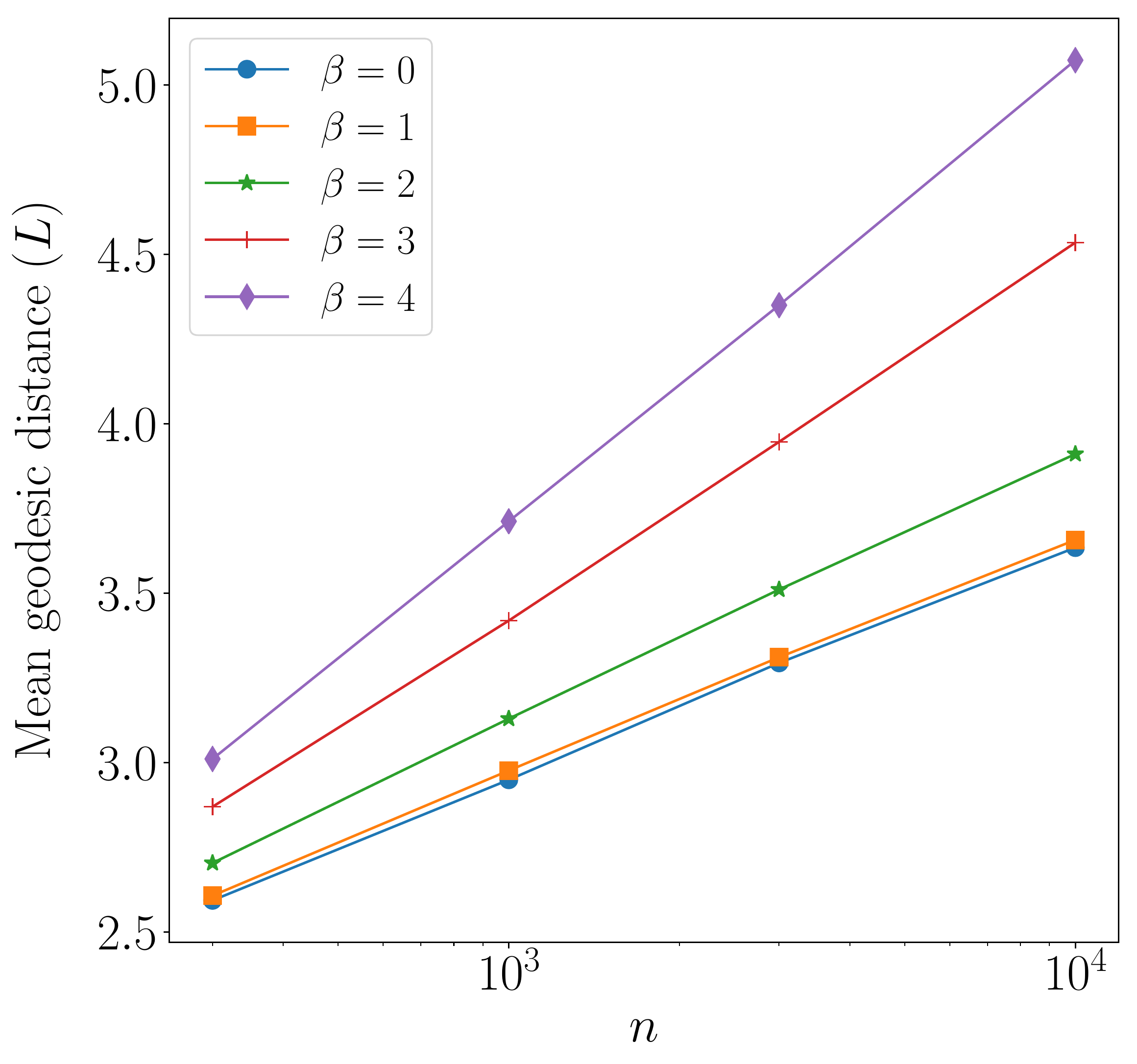}
    \caption{Mean geodesic distance of our SPA networks as a function of the number $n$ of nodes in a network for several values of the spatial decay parameter $\beta$. We show the number of nodes on a logarithmic scale. Each point in the plot represents a mean of the mean geodesic distances over $10$ instantiations of our SPA model.
    }
    \label{fig:PA_geodesic}
\end{figure}

\begin{figure}
    \centering
    \includegraphics[width=0.75\linewidth]{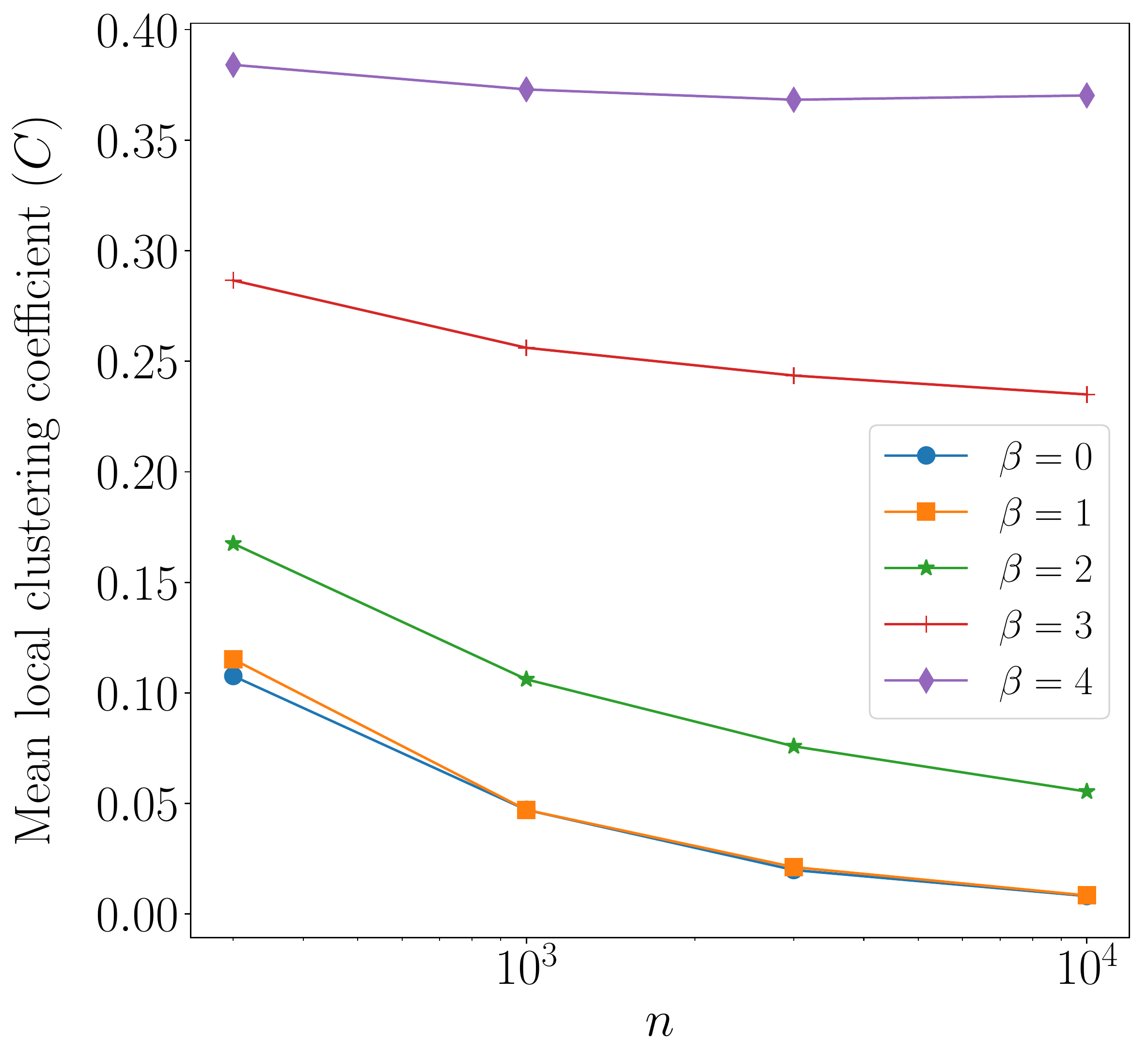}
    \caption{Mean local clustering coefficients of our SPA networks as a function of the number $n$ of nodes in a network for several values of the spatial decay parameter $\beta$. We show the number of nodes on a logarithmic scale. Each point in the plot represents a mean of the mean local clustering coefficients over $10$ instantiations of our SPA model.
    }
    \label{fig:PA_clustering}
\end{figure}

As with the GF model, we calculate the mean local clustering coefficient, mean geodesic distance, mean edge length, and degree assortativity of our networks. 
(See Section \ref{sec:fitness_model} for definitions of these quantities.) In our SPA model, we observe the same general trends in these quantities for progressively larger values of the spatial decay parameter $\beta$ as we observed in the GF model. Specifically, the mean local clustering coefficient, mean geodesic distance, and degree assortativity are larger for progressively larger $\beta$; and the mean edge length is smaller for progressively larger $\beta$.

We examine in more detail how the mean geodesic distance (see Fig.~\ref{fig:PA_geodesic}) and mean local clustering coefficient (see Fig.~\ref{fig:PA_clustering}) depend on the number of nodes for SPA networks. Based on our numerical simulations in Fig.~\ref{fig:PA_geodesic}, for spatial decay parameter values $\beta \in \{ 0, 0.5, 1, 1.5, 2, 2.5, 3, 3.5, 4\}$, the mean geodesic distance seems to increase logarithmically with the number $n$ of nodes. This is consistent with the observations for a 1D SPA in Xulvi-Brunet and I. M. Sokolov \cite{SPA1}, who observed in numerical simulations that mean geodesic distance seems to depend logarithmically on $n$.

In Fig.~\ref{fig:PA_clustering}, we see for $\beta \in \{0,1,2\}$ that the mean local clustering coefficient decays sharply for progressively larger $n$, and it is possible that it may approach $0$ as $n \rightarrow\infty$. However, for $\beta=4$ and $\beta=3$, we do not observe such sharp decay, at least for the examined values of $n$. Our numerical computations suggest the possibility that there is a value $\beta_c$ such that for $\beta > \beta_c$, one obtains a positive mean local clustering coefficient in the limit that $n \rightarrow \infty$.


\section{A Spatial Configuration Model} \label{sec:configuration_model}

We now generalize a configuration model \cite{fosdick, newman2018} to incorporate spatial considerations. Configuration models are among the most important random-graph models, as they are used frequently as reference models (including as null models in community detection \cite{community1, community2}) in network analysis. We envision that spatial analogs of configuration models will be similarly helpful for spatial networks.


\subsection{Description of the model}

In developing a spatial configuration model, we seek to preserve the degree sequence of an input network, while randomizing the adjacencies in the network according to some rule that incorporates a spatial embedding. Specifically, the nodes are embedded in a latent space, and we assign the numbers of stubs to the nodes from the degree sequence of the input network. We then match stubs in a process that resembles the usual one from a non-spatial configuration model \cite{fosdick} (i.e., using a random matching), but instead of selecting edge stubs uniformly at random, we preferentially match stubs that are spatially close to each other.

To undertake this process, we need to make some choices. We must choose how to assign node locations. They can be assigned uniformly at random or according to a different randomization. If the input network is embedded in space and includes node locations, we can also use the existing node locations. We must also choose how to bias stub selection to connect stubs from spatially close nodes. In our spatial configuration model, we use a deterrence function in a similar fashion as in our SPA and GF models. Our procedure is the following:
\begin{enumerate}
    \item For each node, assign a location, which we choose uniformly at random, in $[0, 1] \times [0, 1]$.
    \item To each node $v_l$ with degree $k_l$, assign $k_l$ stubs to it. We denote the number of unmatched stubs for a node $v_l$ at time step $t$ by $u(v_l, t)$.
    \item Choose a stub from step (2) uniformly at random. We use $v_i$ to label its associated node.
    \item Choose a second stub with a probability proportional to $h(r_{i,j})$. That is, select a stub from node $v_j$ with probability
   \begin{equation*}
        p(v_i, v_j, t) = \frac{u(v_j, t)h(r_{i,j})}{\sum_{l \neq i} u(v_l, t)h(r_{i,l})} \,.
    \end{equation*}
    \item Connect the two stubs from steps (3) and (4) to each other with an undirected, unweighted edge. 
    \item Repeat steps (3)--(5) until we have matched all stubs to form edges between the nodes.
\end{enumerate}

Because $r_{i,j}$ is independent of the time step, we can make the above process efficient by calculating the pairwise Euclidean distance between each node pair in a network (there are $O(n^2)$ such pairs to calculate) and store it for reuse in each stub-choosing step. After this, the algorithm takes $O(|E| n)$ time to run, where $E$ denotes the set of edges and $|E|$ is the number of edges.

In formulating a spatial configuration model, one needs to decide whether to allow multi-edges and/or self-edges. Because $h(r_{i,i}) = 0^{-\beta}$ does not have a well-defined value for positive values of $\beta$, we disallow self-edges. However, we do allow multi-edges. As with non-spatial configuration models \cite{fosdick}, choices in the implementation of a spatial configuration model depend on the application and question of interest. One common application of a configuration model is to use it as a null model in community detection \cite{forcechains, community1, community2}, where the exact choice of the null model greatly affects the properties of detected communities. As discussed in \cite{fosdick}, the choice of whether to allow self-edges and/or multi-edges is an important one.

One can also envision many other types of spatial configuration models, and it is worthwhile to study them in future work. For example, it seems interesting to randomize the positions of nodes without rewiring the edges of a network.


\subsection{Computational results}

To illustrate the properties of our spatial configuration model, we start by generating a standard BA network. The seed network for this BA network is a $10$-clique, each new node has $m=5$ stubs, and we grow the network for a total of $T=1000$ time steps (so the final network has $n=1010$ nodes). We then use the degree sequence from this network as the degree sequence for our spatial configuration model. This process gives a single spatial configuration-model network.

For each value of $\beta \in \{0, 0.5, 1, 1.5, 2, 2.5, 3, 3.5, 4\}$, we generate $30$ such networks, and we calculate mean characteristics for these networks. To construct each spatial configuration-model network, we generate a new BA network to create a degree sequence.

\begin{figure}
    \centering
    \includegraphics[width=1.0\linewidth]{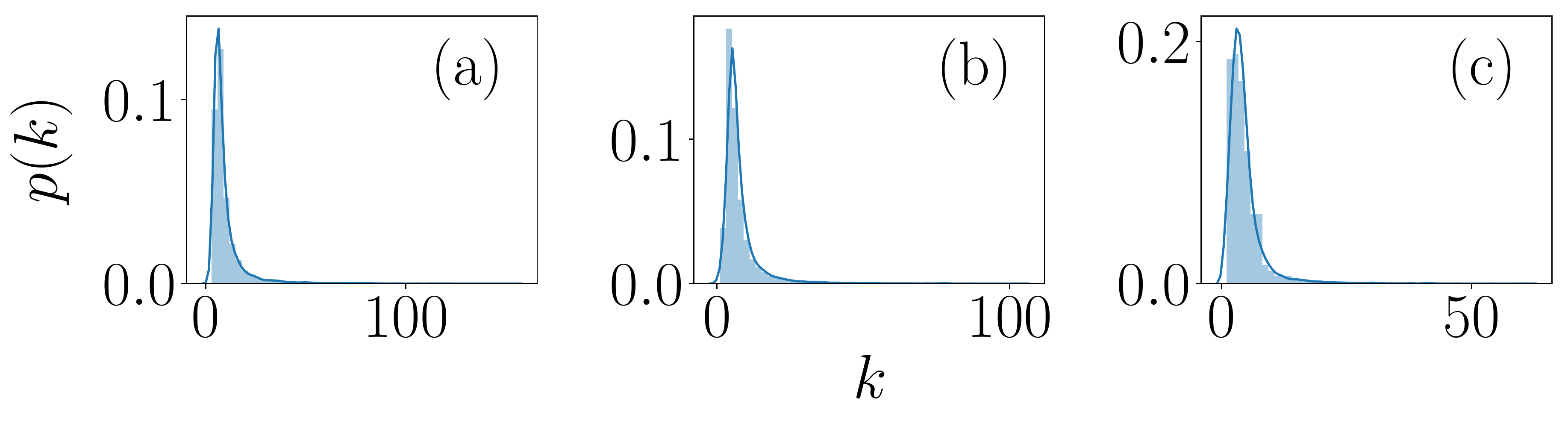}
    \caption{Degree distributions for individual instances of our spatial configuration 
     model, with a degree sequence given by a BA network with $n=1010$ nodes and $m=5$ new edges for each node that we add after the seed. The spatial decay parameter $\beta$ for each panel is (a) 0, (b) 2, and (c) 4.
     }
\end{figure}

\begin{figure}
    \centering
    \includegraphics[width=1.0\linewidth]{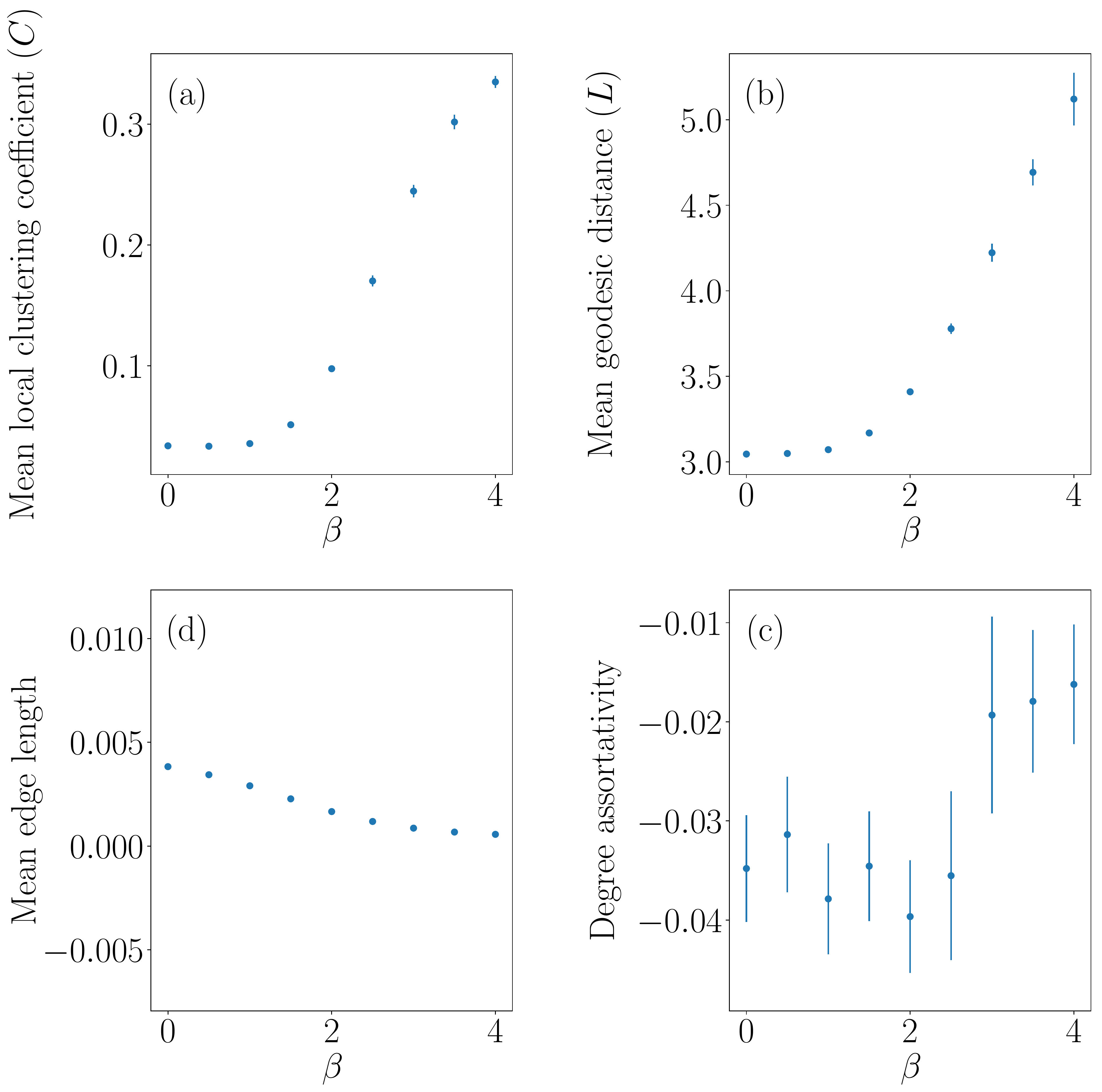}
    \caption{Some characteristics of the networks from our spatial configuration-model networks for various values of the spatial decay parameter $\beta$. For each configuration-model network, we use a degree sequence given by a BA model with $n=1010$ nodes and $m=5$ new edges for each node that we add after the seed. In our computations, we take means over $30$ instantiations (for which we have 30 different networks) of our spatial configuration model.
     We show computations of (a) mean local clustering coefficient, (b) mean geodesic distance, (c) mean edge length, and (d) degree assortativity. The error bars indicate 95\% confidence intervals.
    }
    \label{fig:spatial_configuration_metrics}
\end{figure}

\begin{figure}
    \centering
    \includegraphics[width=1.0\linewidth]{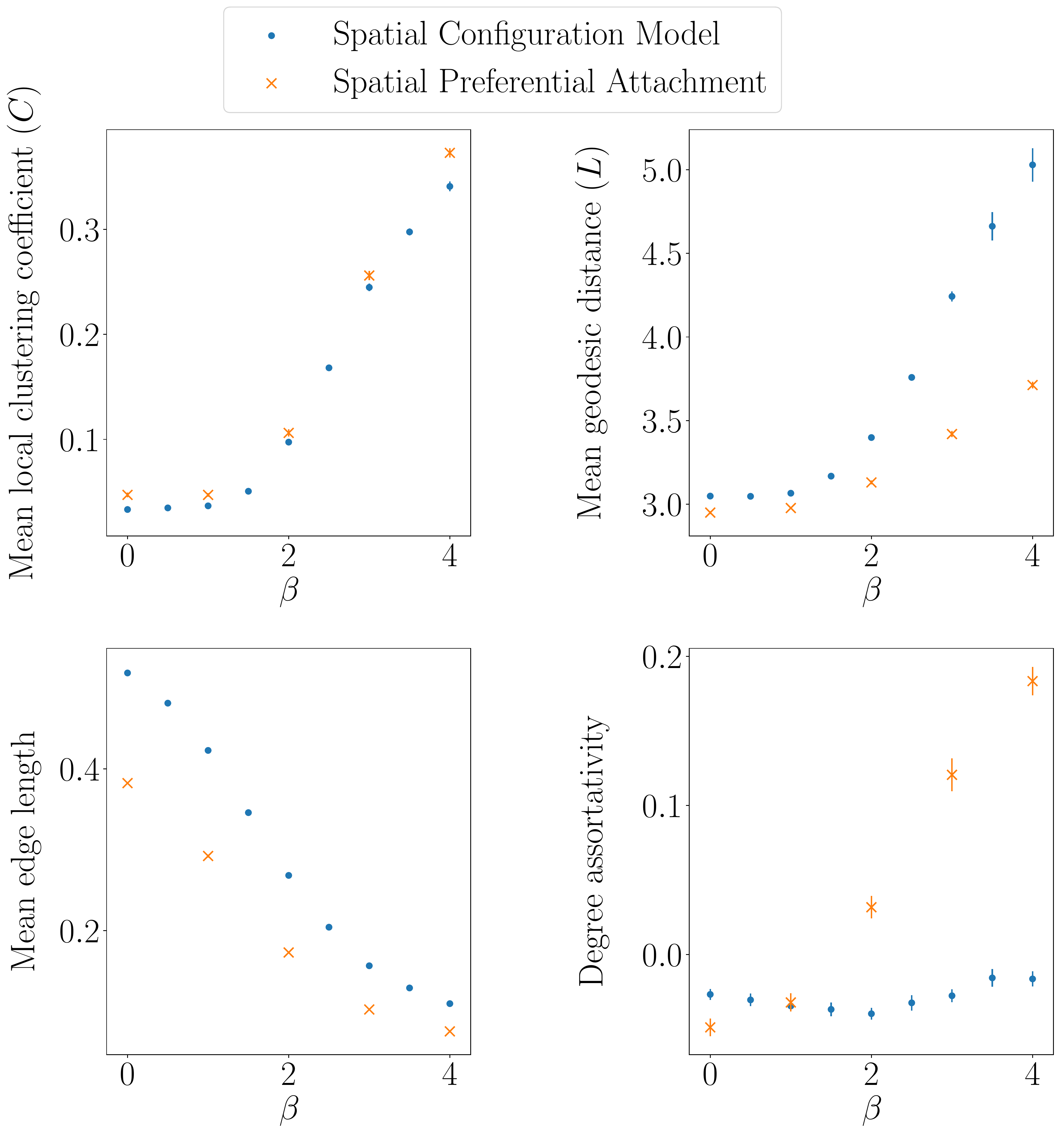}
    \caption{Comparison of the characteristics of our spatial configuration-model networks (blue disks) with characteristics of SPA networks (orange crosses) with $n = 1010$ nodes and $m = 5$ new edges for each node that we add.
     The spatial configuration-model networks are the same networks (which use a degree sequence from a BA network) from Fig.~\ref{fig:spatial_configuration_metrics}. We take all data points as means over $30$ instantiations, and error bars indicate $95\%$ confidence intervals.
}
    \label{fig:spatial_preferential_comparison}
\end{figure}

In our spatial configuration model, we observe (as in our SPA and GF models) for progressively larger values of the spatial decay parameter $\beta$, that the mean local clustering coefficient and mean geodesic distance increase, whereas the mean edge length decreases. The degree assortativity of networks from our spatial configuration model does not appear to have a clear correlation with $\beta$, in contrast to our observations for our SPA and GF models.

For comparison, we include a scatter plot of these characteristics for our spatial configuration model (for which we use degree sequences that we generate from BA networks)  alongside scatter plots from our SPA network from Section \ref{sec:ba-model}. In both the SPA networks and the spatial configuration-model networks that we generate from BA networks, we use a $10$-clique as the seed network, the same total number of nodes ($n=1010$), and the same number of new edges ($m=5$) that we add per node.


\section{Spatial Strength Centrality} \label{sec:spatial_strength}

Our explorations of spatial network models raise an interesting question: Can we quantify the strength of the effects of a spatial embedding and choice of 
deterrence function $h(r)$? We have observed that larger values of the spatial decay parameter $\beta$ lead to more prominent spatial effects on network topology. However, it is desirable to be more systematic about our analysis of spatial networks. For example, it is important to compare different choices of the deterrence function $h(r)$, different sizes and dimensions of the ambient space, and different distributions of nodes in space. Therefore, we define a centrality measure for spatial networks that we call \emph{spatial strength centrality}, and we study it in several synthetic and empirical spatial networks.


\subsection{Definition and description of spatial strength centrality}

To develop a notion of centrality for spatial networks, we proceed as follows. Let $N(v_i)$ be the neighborhood of $v_i$ (i.e., the set of nodes to which node $v_i$ is adjacent). We calculate a normalized mean edge distance --- which we take to be Euclidean for concreteness, but one can also consider other metrics --- from node $v_i$ to each node in its neighborhood. For a node $v_i$ with at least one incident edge, this distance is
\begin{align}
    L(v_i) = \frac{\sum_{v_j \in N(v_i)}r_{i,j}}{k_i} \frac{1}{{\langle L \rangle}}\,,
\label{eq:v_edge_length}
\end{align}
where
\begin{align}
    \langle L \rangle = \frac{\sum_{(v_i, v_j) \in E}r_{i,j}}{n \langle k \rangle}
\end{align}
and $r_{i,j}$ is the Euclidean distance between nodes $v_i$ and $v_j$. The left fraction in $L(v_i)$ gives the mean edge length of $v_i$; we then normalize it by $\langle L \rangle$, the mean edge length in the network. We now calculate the mean neighbor degree of each node $v_i$ and normalize it by the mean degree in the network. That is,
\begin{equation}
    K(v_i) = \frac{\sum_{v_j \in N(v_i)} k_j}{k_i} \frac{1}{\langle k \rangle}\,.
\end{equation}
We then combine the above two quantities to calculate the spatial strength centrality
\begin{equation} \label{eq:spatial_strength}
    S(v_i) := \frac{1}{L(v_i)K(v_i)}
\end{equation}
of each node $v_i$ with at least one incident edge.

Our motivation in defining $S(v_i)$ is to capture a notion of whether nodes are adjacent to each other because they are spatially close or because they are adjacent to each other for a topological reason. Heuristically, we reward a node for being adjacent to spatially close nodes, and we penalize it for being adjacent to node with large degree (which, in some contexts, are ``hubs'').

As an example, consider a network of flights between airports. Suppose that a small airport $v_i$ is adjacent only to a hub $v_j$, which is also far away from it geographically. Node $v_i$ has a small spatial strength $S(v_i)$ because its edge to $v_j$ does not arise from the fact that it is nearby, but instead because $v_j$ is a hub. We are able to capture this idea with $S(v_i)$ because both $L(v_i)$ and $K(v_i)$ are large, as the one edge of node $v_i$ is a long edge and its neighbor has large degree. Therefore, from \eqref{eq:spatial_strength}, we see that $S(v_i)$ is small. In Section \ref{synth}, we explore a toy example (specifically, a hub-and-spoke network)
of this situation in more detail.

Conversely, consider a granular network \cite{papa2018}. In such a network, nodes are adjacent if they are touching (or at least sufficiently close to be construed as touching). Because of physical constraints, the number of edges that are attached to a node is constrained to be small. Moreover, a node $v_j$ in a granular network has short edges, so $L(v_j)$ is small. Because every node in the neighborhood of $v_j$ has small degree, it follows that $K(v_j)$ is small. Therefore, from \eqref{eq:spatial_strength}, we expect $S(v_j)$ to be larger in this example than $S(v_j)$ in our example of flights between airports.

It can also be informative to calculate a network's mean spatial strength centrality
\begin{equation}
    \langle S \rangle = \frac{\sum_i S(v_i)}{n}\,.
\end{equation}
In the previous examples, for instance, we expect a granular network to have a larger value of $\langle S \rangle$ than the network of flights between airports, because the former's network topology is subject to more stringent constraints.

There are several important considerations for calculating mean spatial strength:
\begin{enumerate}
    \item We have normalized all quantities, so we expect to be able to meaningfully compare the values of $\langle S \rangle$ for different types of networks, including ones with different sizes and spatial embeddings. However, $\langle S \rangle$ is unbounded (in particular, it is not confined to values between $0$ and $1$), so we need to be careful about interpreting its values and comparisons of these values.
    \item The mean spatial strength $\langle S \rangle$ is nonnegative.
    \item We normalized $K(v_i)$ by the mean degree $\langle k \rangle$, rather than by the largest degree $k_{\mathrm{max}}$ in a network, so it is not guaranteed to lie between $0$ and $1$.
    \item Our formulas for $L(v_i)$ and $S(v_i)$ are not well-defined for nodes that have no edges; we take these quantities to be $0$ in these cases.
\end{enumerate}


\subsection{Computation of spatial strength on network models}\label{sec:computed_ss}

\begin{figure}
    \centering
    \includegraphics[width=1.0\linewidth]{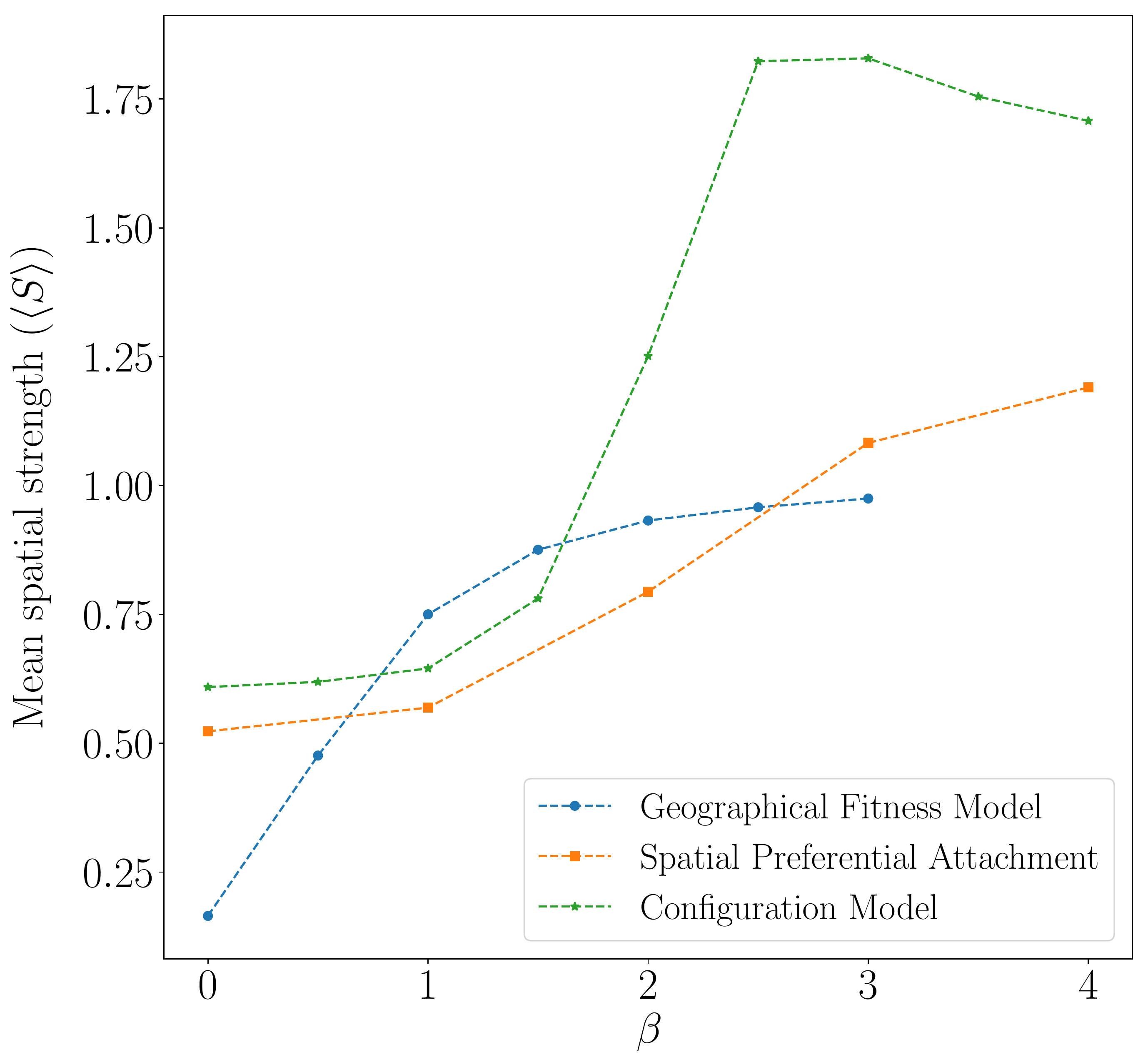}
    \caption{Mean spatial strength centralities for our GF model, our SPA model, and our spatial configuration model for various values of the spatial decay parameter $\beta$. For a given value of $\beta$, each point represents a mean over $20$ realizations of a model. Our GF-model networks have $n=500$ nodes, and our SPA and spatial-configuration-model networks have $n=1010$ nodes.
    }
    \label{fig:spatial_model_strength}
\end{figure}

As initial test cases, we examine spatial strength centralities in our GF model, SPA model, and spatial configuration model. We expect the mean spatial strength $\langle S \rangle$ to become progressively larger for progressively larger values of the spatial decay parameter $\beta$ (see Fig.~\ref{fig:spatial_model_strength}).

Interestingly, although there is generally a positive correlation between $\beta$ and $\langle S \rangle$, it is not a linear relationship. The spatial configuration and SPA models have S-shaped curves, and the GF model has an initially rapid increase of $\langle S \rangle$ with $\beta$ before tapering off.

Contrary to our expectations (which were for $\langle S \rangle$ to tend to increase with $\beta$), the spatial configuration model has a peak in the mean spatial strength centrality at about $\beta = 3$. For progressively larger $\beta$, the mean edge length of a network decreases, in turn decreasing the mean spatial strength centrality (because $L(v_i)$ increases as $\langle L \rangle$ decreases, as we can see from \eqref{eq:v_edge_length}, so $S(v_i)$ decreases as $\langle L \rangle$ decreases). This sensitivity to mean edge length is a potential weakness in our centrality measure. In Section \ref{sec:alternate_formations}, we discuss possible ways to address this issue.

\begin{figure}
    \centering
    \includegraphics[width=1.0\linewidth]{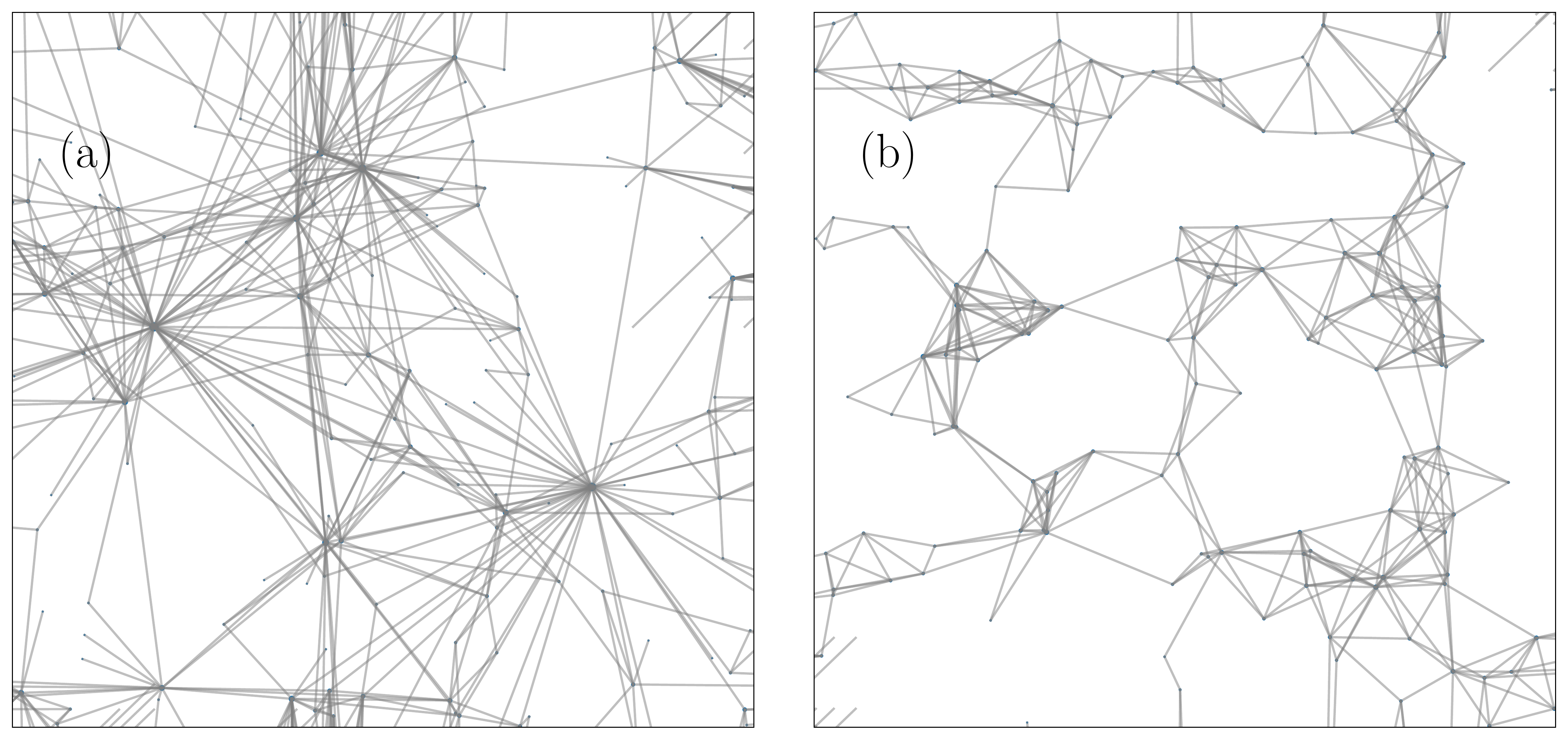}
    \caption{Mean spatial strength centralities of example networks from individual instantiations of the GF model with $n=150$ nodes, which we place in $[0,1] \times [0,1]$ (with periodic boundary conditions) according to their assigned coordinates. 
    We show examples with (a) $\beta = 0.5$ and $\langle S \rangle \approx 0.786$ and (b) $\beta = 3$ and $\langle S \rangle \approx 1.007$.
    }
    \label{fig:network_visualizations}
\end{figure}


\subsection{Examination of spatial strength centrality on simple synthetic networks}\label{synth}

To develop intuition about spatial strength centrality, we consider several simple types of synthetic networks. We start with our GF model, for which we illustrate instantiations of different mean spatial strength centralities in Fig.~\ref{fig:network_visualizations}. 

\begin{figure}
    \centering
    \includegraphics[width=1.0\linewidth]{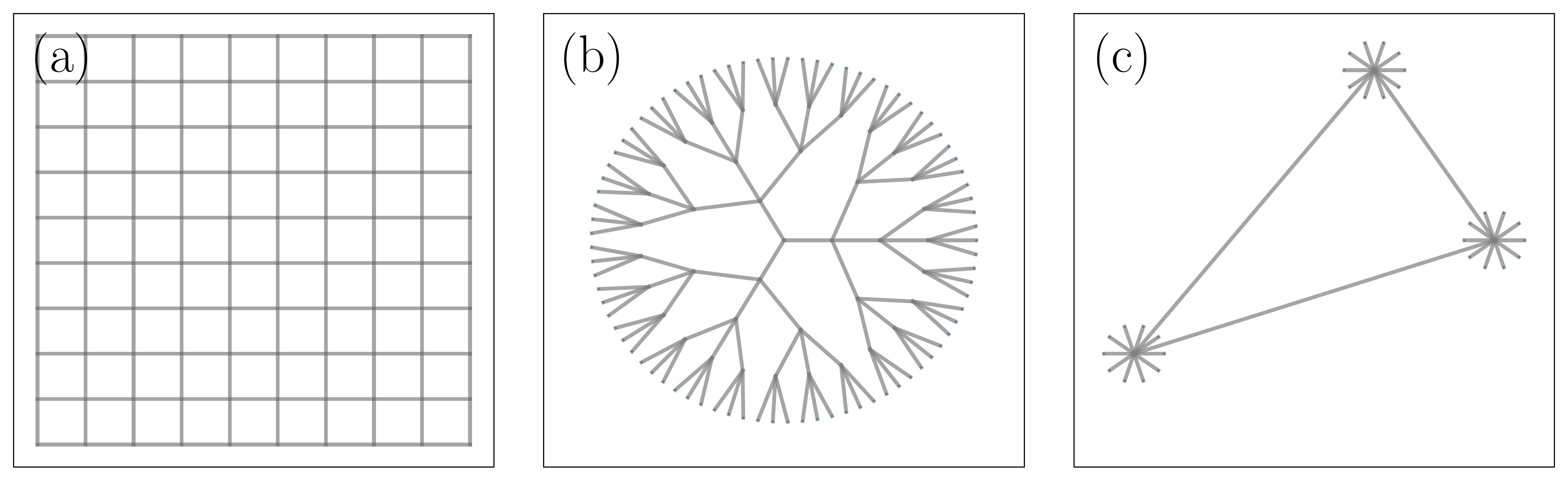}
    \caption{Example networks with the indicated values of mean spatial strength centrality. The depicted networks are (a) a lattice network, $\langle S \rangle \approx 0.999$, (b) a spatially-embedded Cayley tree $\langle S \rangle \approx 0.687$, and (c) a hub-and-spoke example $\langle S \rangle \approx 0.300$.
    }
    \label{fig:toy_network_examples}
\end{figure}

\begin{figure}
    \centering
    \includegraphics[width=1.0\linewidth]{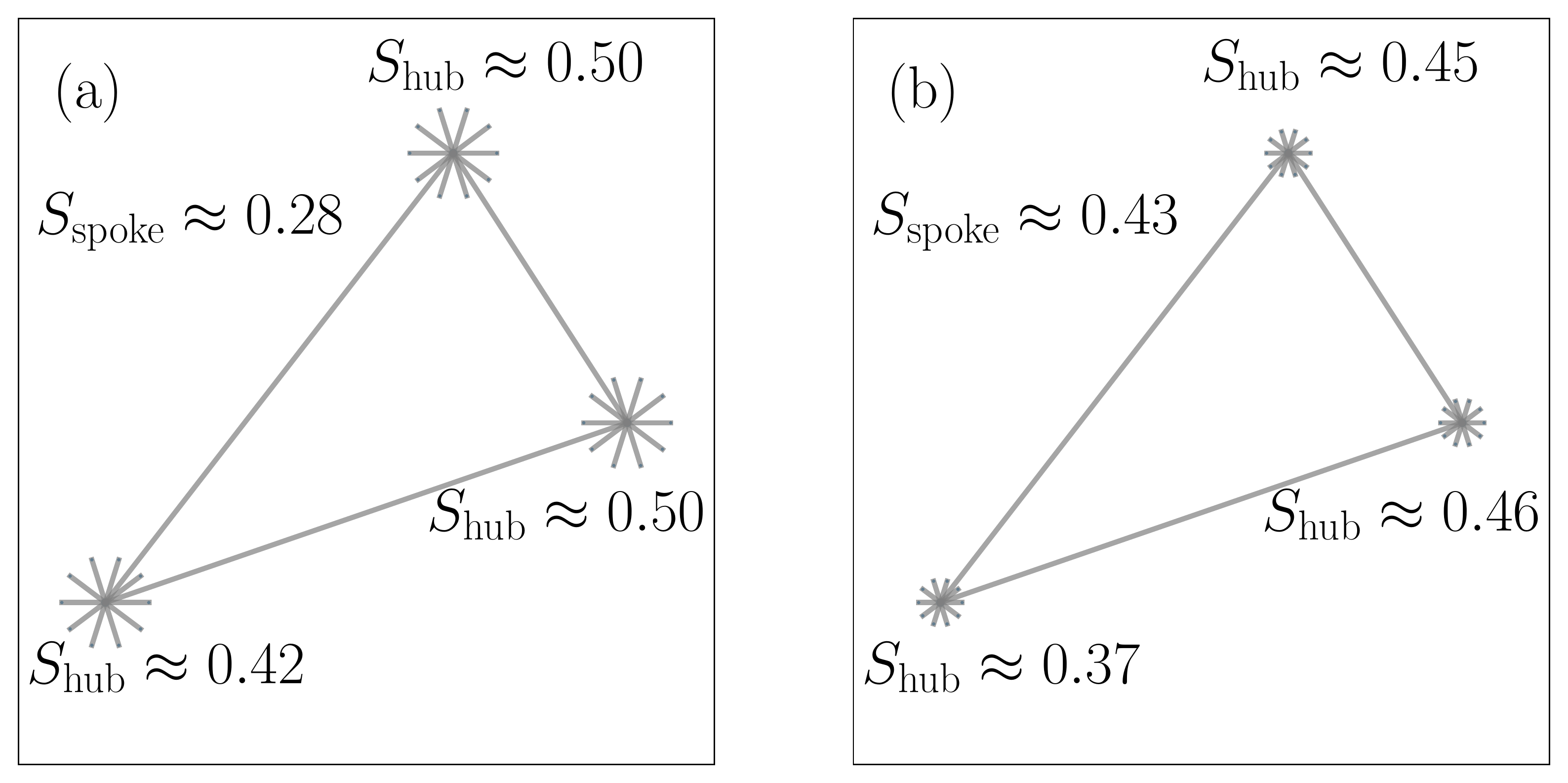}
    \caption{These small hub-and-spoke networks give a toy example that illustrates an idea of potential relevance to a network of flights between airports. For each network, we give the spatial strength centralities of each hub and of each spoke. In each of the two examples, all spoke nodes have the same spatial strength centrality. Panel (a) has $\langle S \rangle \approx 0.300$, and panel (b) has $\langle S \rangle \approx 0.430$. In panel (b), the hub--spoke edges are half the length of such edges in panel (a). Notably, the values of $S_{\text{hub}}$ are smaller in the right network.
    }
    \label{fig:hub_spoke_example}
\end{figure}

\begin{figure}
    \centering
    \includegraphics[width=0.4\linewidth]{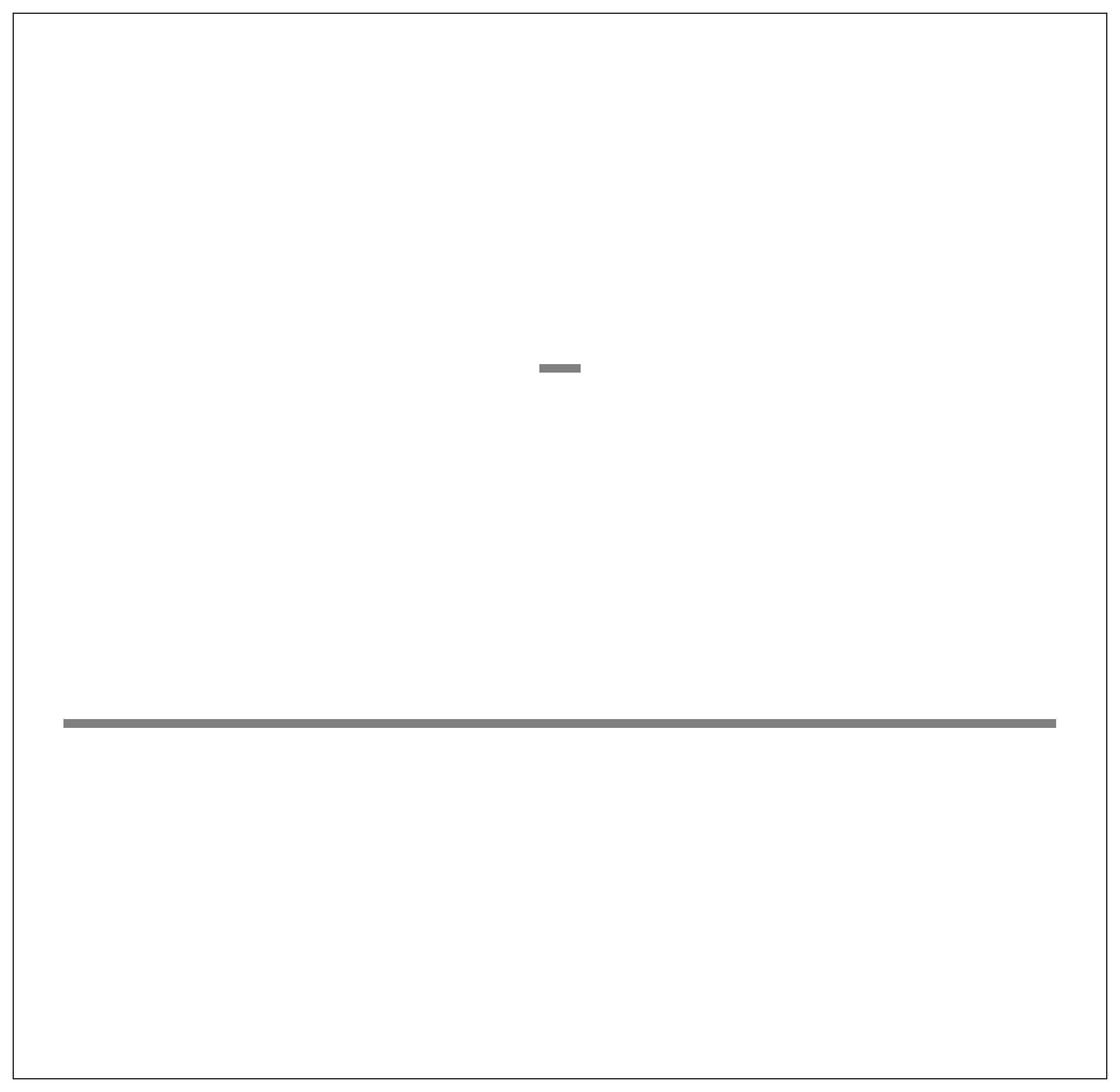}
    \caption{This example of a two-scale network with two pairs of nodes has $\langle S \rangle \approx 8.009$ and illustrates that the mean spatial strength centrality $\langle S \rangle$ can be arbitrarily large.
    }
    \label{fig:breaking_example}
\end{figure}

We also consider the following types of networks:
\begin{itemize}
\item{\textbf{Square lattice.} The lattice network has $x$ columns and $y$ rows, giving it a total of $n = xy$ nodes. We space these nodes evenly in $[0,1] \times [0,1]$; each node is adjacent to the nodes that are immediately north, south, east, and west of it. See the left plot of Fig.~\ref{fig:toy_network_examples}. When $x=y$, the mean spatial strength tends towards $1$ as $n \rightarrow \infty$.

\item{\textbf{Spatially-embedded Cayley tree.} This network has $b$ branches and $l$ layers. We start with one central node (layer $0$), and each node in layer $l$ is adjacent to $b$ nodes in layer $l+1$, whose nodes are equally spaced in a circle of radius $l+1$. See the center plot of Fig.~\ref{fig:toy_network_examples}. 
}
\item{\textbf{Hub-and-spoke example.} This example has three ``hub'' nodes that are each spaced relatively far away from its $10$ ``spoke'' nodes, which occur in a circle around it. See Fig.~\ref{fig:hub_spoke_example}. This example demonstrates that a network with long-range connections can have a small mean spatial strength centrality. We also observe that hub-and-spoke networks with shorter hub--spoke edges tend to have larger mean spatial strength centralities. (Compare the left and right panels of Fig.~\ref{fig:hub_spoke_example}.)

One may perhaps expect that hubs tend to have larger spatial strength centralities than spokes, as hubs $v_i$ have smaller values of $K(v_i)$. However, when we decrease the hub--spoke edge length in Fig.~\ref{fig:hub_spoke_example} [compare panel (b) to panel (a)], we observe that $S_{\text{spoke}} > S_{\text{hub}}$ for the bottom-left hub. If we decrease this edge length further, $S_{\text{spoke}} > S_{\text{hub}}$ for all hubs. This occurs because the value of $\langle L \rangle$ of a hub-and-spoke network decreases as we decrease the hub--spoke edge length, so the value of $L(v_i) $ for hubs $v_i$ increases relative to $\langle L \rangle$. This example demonstrates how spatial strength centrality may point to interesting properties even in a simple toy network. It also suggests that investigating the spatial strength centrality of individual nodes may often be more insightful than considering only the mean spatial strength centrality of a network.
  }
  }
\item{\textbf{Two-scale example.} Exploiting the definition of spatial strength centrality, we construct a network that consists of two pairs of nodes, where the nodes of each pair are adjacent to each other (see Fig.~\ref{fig:breaking_example}). By making one edge short and the other edge arbitrarily long, the mean spatial strength centrality approaches infinity as the second edge becomes arbitrarily long. This example demonstrates that there exist networks with arbitrarily large values of mean spatial strength centrality. 
}
\end{itemize}

\subsection{Examination of spatial strength centrality on empirical and synthetic data sets}\label{data}

\begin{figure*}
    \centering
    \includegraphics[width=1.0\linewidth]{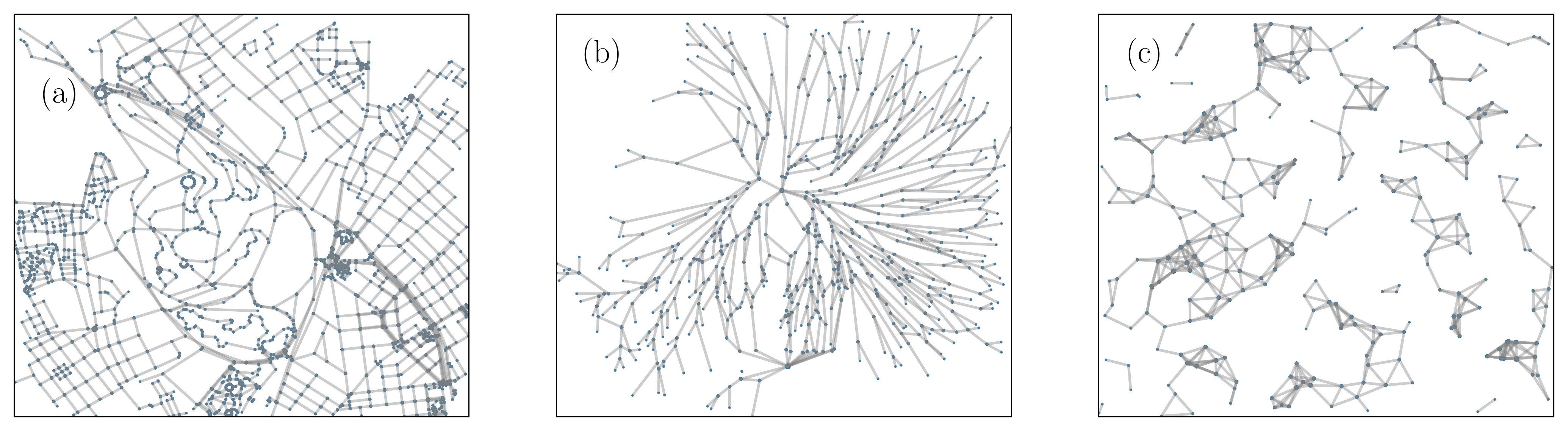}
    \caption{Example networks and mean spatial strength centralities of empirical networks and an RGG. We show (a) a road network from Tunis, Africa with $1731$ nodes and $\langle S \rangle \approx 2.339$; (b) a fungal network of type ``Pv\textunderscore M\textunderscore I+4R\textunderscore U\textunderscore N\textunderscore 21d\textunderscore 4'' (see \cite{fungal_data}) with $641$ nodes and $\langle S \rangle \approx 2.279$; and (c) an RGG with $n = 500$ nodes, a connection radius of $r_c = 0.07$, and $\langle S \rangle \approx 1.228$.
    }
    \label{fig:data_network_examples}
\end{figure*}

We now examine spatial strength centrality in several empirical networks, as well as on an RGG (which we defined in Sec.~\ref{sec:fitness_model}).
The data sets for fungal networks are from \cite{fungal_data} (with $270$ networks with between $68$ and $2742$ nodes and a mean of $819$ nodes), and the data sets for city road networks are from \cite{road_data} (with $101$ networks with between $42$ and $3871$ nodes and a mean of $874$ nodes). We show some example networks and their mean spatial strength centralities in Fig.~\ref{fig:data_network_examples}.

None of the model networks that we have explored in depth in the present paper have yielded a mean spatial strength centrality that is larger than $2$. However, see Fig.~\ref{fig:breaking_example} for an illustration that mean spatial strength centrality can be arbitrarily large. Interestingly, many of the networks in both the city road and fungal data sets have a mean spatial strength centrality that is larger than $2$. Therefore, there are structural features in spatial networks beyond the ones in the main models in this paper.

In the examined networks with the largest values of $\langle S \rangle$, we observe regions of space in which nodes are very close together.
For example, in a city road network (see the left side of Fig.~\ref{fig:data_network_examples}), multiple nodes (i.e., street intersections) can occur along curves in a road. Such nodes tend to have short edges between them, and these networks thus include edges at multiple spatial scales. (Recent developments in topological data analysis \cite{feng2019,feng2020} have examined such multiscale phenomena, providing a complementary perspective to that of the present paper.) As we saw in the example in Fig.~\ref{fig:breaking_example}, having both very long edges and very short edges can lead to a large mean spatial strength centrality.

\begin{figure}
    \centering
    \includegraphics[width=1.0\linewidth]{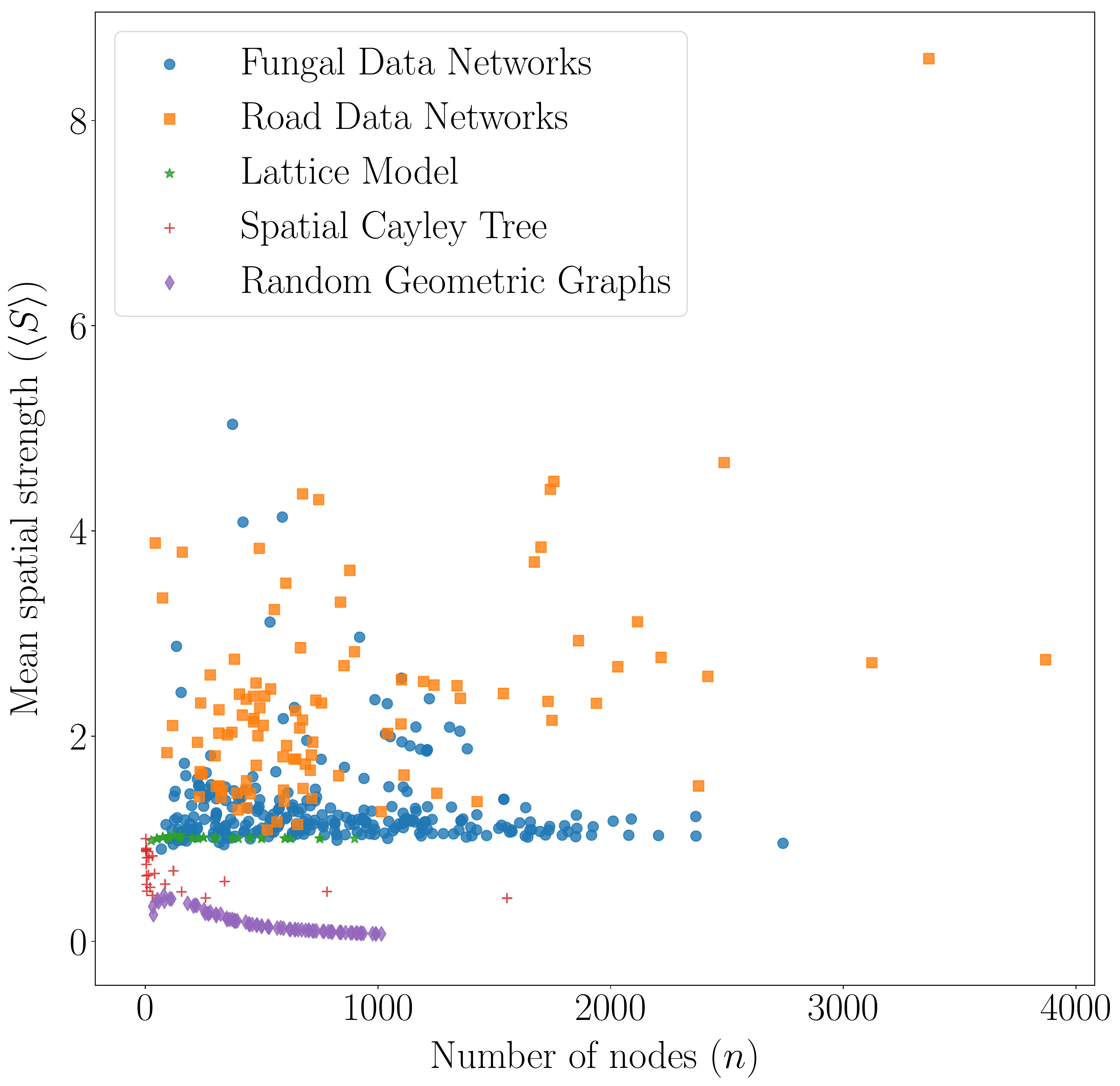}
    \caption{Scatter plot of mean spatial strength centrality versus the number of nodes in several synthetic and empirical networks. We use fungal network data from \cite{fungal_data}, road network data from \cite{road_data}, and RGGs with a connection radius of $r_c = 0.07$. The spatial Cayley trees in this plot have from $1$ to $6$ branches, and their depths range from $1$ to $4$. The lattice networks have $5$, $10$, $15$, $20$, $25$, or $30$ rows and columns (and any combination of rows and columns of those sizes); in the plot, these form an almost horizontal line at a mean spatial strength centrality of about $1$.
    }
    \label{fig:spatial_distributions}
\end{figure}

In Fig.~\ref{fig:spatial_distributions}, we show a scatter plot of many of these networks and their mean spatial strength centralities.


\subsection{Alternate formulations of spatial strength centrality}\label{sec:alternate_formations}

As we saw in our computations of spatial strength centrality, it is sometimes able to capture some aspects of how spatial embeddings influence network topology, but it is not always successful at doing so (e.g., the fact that $\langle S \rangle$ is not an increasing function of $\beta$ for the spatial-configuration-model networks). To explore these issues further, it is worth considering the following ideas:
\begin{itemize}
    \item{\textbf{Calculating spatial strength centralities for edges.} In Eq.~\eqref{eq:spatial_strength}, we defined spatial strength centrality as a combination of mean neighbor degree and the mean edge length of a node. One can examine the ``spatial strength centrality'' of edges, instead of nodes. For example, for an edge $(i, j)$, one can sum the degrees of $v_i$ and $v_j$ and then divide by the length of the edge $r_{i,j}$. }
    \item{\textbf{Normalization of mean edge length.} We noted (see Section \ref{sec:computed_ss}) that $\langle S \rangle$ is sensitive to mean edge length. This sensitivity occurs because $L(v)$ increases and $S(v)$ decreases as $\langle L \rangle$ decreases. 
  As an alternative, it may be useful to normalize $L(v)$ by the maximum pairwise edge length in a network, rather than by the mean edge length. We chose the latter to be able to compare networks of different sizes. An alternate way to allow such a comparison is to normalize $L(v)$ by the geographical diameter of a network, as measured by the maximum distance between two nodes of a network.
  }
    \item{\textbf{Comparison of a spatial network to null models.} In developing and calculating spatial strength centrality, we seek to examine how a network's spatial embedding affects its structure and characteristics. However, instead of only calculating a centrality measure, it may be desirable to compare a network to a spatial null model to determine how a spatial embedding affects the adjacency matrix (and hence structure) of the network. Our spatial configuration model may be useful as a null model for such purposes.}
\end{itemize}


\section{Conclusions and Discussion} \label{sec:discussion}

We have developed and examined a straightforward method for generalizing generative models of networks to incorporate spatial information by using a deterrence function $h(r)$ that decays with the distance $r$ between nodes to adjust the probability that an edge occurs between a pair of nodes. For concreteness, we used Euclidean distance and the power-law decay rate $h(r) = r^{-\beta}$, but our formulation allows one to make diverse choices of both metric and decay function. 

One illustration of how to augment existing network models, rather than define new spatial network models from scratch, is with our formulation of a spatial configuration model. We also extended a geographical fitness (GF) model with a deterrence function $h(r)$ and studied an spatial preferential attachment (SPA) model that uses this deterrence function. We examined the properties of these models and compared them to random geometric graphs and empirical spatial networks from two disparate applications. To examine the structure of spatial networks more deeply --- and, in particular, to try to separate the effects of spatial embeddings and other influences on network architecture --- we defined a spatial strength centrality, which allowed us to estimate how strongly the effects of a network's ambient space (in which its nodes are embedded) affects observed network topology. We then examined spatial strength centrality on several toy networks and compared it on a diverse set of synthetic and empirical spatial networks.

Spatial networks have diverse uses in the modeling of networks from empirical data, and the models that we have examined in the present paper should help in such efforts. We anticipate that further exploration of spatial null models (e.g., using our spatial configuration model and generalizations of it) will be particularly insightful, as they provide baselines for comparisons with empirical data. To explore the diverse effects of spatial embeddings (and other effects of space) on network topology, it is also important to further analyze deterrence functions $h(r)$ and a variety of notions of spatial strength centrality.


\section*{Acknowledgements}

We thank Heather Brooks and two anonymous referees for helpful discussions.



%

%
\end{document}